\documentclass[%
 reprint,
 showpacs,preprintnumbers,
 nofootinbib,
 amsmath,amssymb,
 aps,
 prd,
 floatfix,
]{revtex4-1}
\usepackage{graphicx}

\usepackage{amssymb,amsmath,amstext,amsthm,amsfonts}
\usepackage{subfigure}
\usepackage[utf8]{inputenc}
\usepackage{float}
\usepackage{url}
\usepackage{todonotes}
\usepackage{tikz}
\usepackage[percent]{overpic}

\begin{document}
	\title{Exploring Nuclear Effects in Neutrino-Nucleus Interactions Using Measurements of Transverse Kinematic Imbalance from T2K and MINERvA}
	
	\author{S. Dolan}
		\email[Contact e-mail: ]{Stephen.Dolan@llr.in2p3.fr}
	\affiliation{Ecole Polytechnique, IN2P3-CNRS, Laboratoire Leprince-Ringuet, Palaiseau 91120, France }
	\affiliation{IRFU, CEA Saclay, Gif-sur-Yvette 91190, France}

\begin{abstract}
\noindent Predictions from widely-used neutrino-nucleus interaction event generators are compared to measurements of transverse kinematic imbalances, made by both the T2K and MINERvA experiments, to allow a joint characterisation of the nuclear physics processes responsible for some of the largest uncertainties in measurements of neutrino oscillations. The role of nucleon-nucleon correlations, initial state nucleon Fermi motion and hadronic re-interactions inside the nuclear medium are explored and areas requiring more theoretical input are identified. 
\end{abstract}

\maketitle
\section{Introduction}	
\label{sec:intro}

In order to determine the neutrino mass hierarchy and the extent of CP-violation generated in the neutrino sector using long-baseline neutrino oscillation experiments, the systematic uncertainty from future measurements must be constrained to within a few percent. One of the main difficulties in achieving such precision stems from our naivety of few-GeV neutrino-nucleus interactions. The associated systematic uncertainties are already dominant in some current measurements of oscillation parameters~\cite{Abe:2017vif, Duffy:2016phs} and will soon become the principle limitation if an improved understanding cannot be achieved. In particular, to make precision measurements of neutrino oscillations it is essential to be able to reconstruct the incoming neutrino's energy from the final state of its interaction, but current methods of doing so are severely obfuscated by poorly-understood nuclear physics processes (‘nuclear effects’).

For the purposes of this manuscript, these nuclear effects can broadly be split into three categories: the initial state motion of nucleons inside a nucleus (Fermi motion) with momenta up to $p_F\sim230$~MeV/c in carbon; nucleon correlation effects, which can sometimes lead to two nucleon, or `two particle two hole' (2p2h) final states; and final state re-interactions (FSI) of hadrons inside the nuclear medium which can both alter the kinematics of the final state and stimulate nuclear absorption or emission (of nucleons or mesons) thereby altering the final state particle content of the interaction. In both the `kinematic method' of neutrino energy reconstruction used by T2K~\cite{Abe:2017vif} and the `calorimetric method' used by NOvA~\cite{NOvA:2018gge} the impact of these nuclear effects are corrected using a neutrino interaction simulation, but the models used to do so are often approximate and subject to very large uncertainties~\cite{Alvarez-Ruso:2017oui}.



These nuclear effects can be better understood by analysis of neutrino-nucleus scattering data. The majority of available measurements report only observables formed from the kinematics of the outgoing lepton, but in such results the wide-band nature of neutrino beams means that the incoming neutrino energy is typically poorly known and therefore the true properties of the interaction (e.g. energy-momentum transfer) are obscured. As a consequence it has been difficult to distinguish the contributions of nuclear effects. For example, the impact of 2p2h is often predicted to be spread over a wide range of lepton kinematics and is often degenerate with plausible neutrino flux-normalisation shifts or alterations to the axial nucleon form-factor~\cite{Gallmeister:2016dnq, Nieves:2011yp}. However, recent analyses by both the T2K~\cite{Abe:2018pwo} and MINVERvA~\cite{Lu:2018stk} experiments, which measure the outgoing lepton and hadron kinematics and consider imbalances on the plane transverse to the incoming neutrino, are able to offer a much more direct probe of nuclear effects~\cite{Lu:2015tcr}. Not only does this projection greatly mitigate the impact of the unknown neutrino energy, but it also allows a measurement of the `missing momentum' that has allowed an interesting probe of nuclear effects in electron-nucleus scattering (e.g.~\cite{Udias:2001tc}) in the neutrino case. The T2K measurement has already been used to characterise 2p2h strength within the GiBUU theory framework~\cite{Dolan:2018sbb,Buss:2011mx}.

In this manuscript these T2K and MINERvA results are analysed together in the context of the models available in some of the most widely used neutrino-nucleus interaction generators. Since the two experiments operate using quite different energy neutrino beams (peaked at around 0.6~GeV for T2K and 3.5~GeV for MINERvA), the simultaneous analysis of both results offers a probe of nuclear effects spanning the relevant energies of all current and planned accelerator-based neutrino oscillation experiments. By exploring the concurrences and discrepancies in model comparisons to the two sets of results, it is possible to assess the robustness of the nuclear models responsible for some of the largest systematic uncertainties in neutrino oscillation analyses, thereby offering both an opportunity to reduce these uncertainties and to identify areas where more theoretical input is required.  

\section{Measurements}	
\label{sec:measurements}

Both T2K and MINERvA measure muon-neutrino `charged-current quasi-elastic like' (CCQE-like) interactions on a hydrocarbon target, defined as those leaving one muon, at least one proton and no mesons in the final state. Kinematic correlations are then analysed between the muon and highest momentum proton. One such set characterise the aforementioned kinematic imbalance in the plane transverse to the incoming neutrino~\cite{Lu:2015tcr}. They are defined by: 
\begin{eqnarray}
\delta p_T &=& |\mathbf{\delta p}_T| = \left| \mathbf{p}_T^\mu + \mathbf{p}_T^p \right|,   \\
\delta \alpha_T &=& {\rm arccos}\left(- \frac{\mathbf{p}_T^\mu \cdot \mathbf{\delta p}_T}{p_T^\mu \delta p_T} \right), \ \\
\delta \phi_T &=& {\rm arccos}\left( - \frac{\mathbf{p}_T^\mu \cdot \mathbf{p}_T^p}{p_T^\mu p_T^p}\right).
\end{eqnarray}
Here $\mathbf{p}_T^\mu$ and $\mathbf{p}_T^p$ are the momentum of the outgoing muon and highest momentum proton in the plane transverse to the incoming neutrino, respectively. 

In addition to the STV, MINERvA also measure the `reconstructed neutron momentum' ($p_n$), proposed in~\cite{Furmanski:2016wqo}, and is equivalent to the magnitude of the total momentum imbalance ($\delta p = (\delta p_L,\mathbf{\delta p}_T)$, where $\delta p_L$ is the longitudinal momentum imbalance) calculated under the assumption that there is a single proton in the final state such that:

\begin{eqnarray}
\delta p_L&=&\frac{1}{2}R-\frac{m_N^2+\delta p_T^2}{2R}\text{, where}\\
R&=&m_N+p_L^\mu+p_L^p-E^\mu-E^p\text{, and}\\
m_N&=&m_T-m_N+E_b.
\end{eqnarray}

where $p_L^\mu$/$E^\mu$ and $p_L^p$/$E^p$ are the muon and proton longitudinal-momenta/energy; $m_T$ is the nuclear target mass; $E_b$ is the binding energy (taken as 27.13 MeV for Carbon~\cite{Furmanski:2016wqo}); such that $m_N$ is the mass of the assumed nuclear remnant. In the case where these assumptions are true (i.e. for pure CCQE interactions with no FSI) then $p_n$ is the momentum of the initial-state target nucleon. Under these same assumptions $\delta p_T$ is the transverse projection of $p_n$. 

It is worth noting that T2K also measures other muon-proton correlations (such as the proton kinematics in bins of muon momentum and angle and the `inferred kinematic imbalance' between them) which may prove to be an interesting and complementary probe of nuclear effects for future model comparisons. 

Both T2K and MINERvA measure fiducial cross sections, where the measured cross sections are constrained to specific kinematic phase spaces, which are defined in Tab.~\ref{tab:phaseSpace}. For T2K, interactions where the highest momentum proton falls outside of the phase-space constraints are not considered, whilst for MINERvA the event is still considered if another proton falls within the constraints (and the observables are then calculated using its kinematics). 

\begin{center} 
\begin{table}[htbp!]
\footnotesize
\begin{tabular}{ |l|c|c|c|c| } 
 \hline
Analysis & $p_p$ & $cos\theta_p$ & $p_\mu$ & $cos\theta_\mu$  \\
 \hline
T2K &  $0.45-1.0$~GeV & $>0.4$ & $>250$~MeV & $>-0.6$ \\
MINERvA &  $0.45-1.2$~GeV & $>0.342$ & $1.5-10$~GeV & $>0.940$ \\
\hline
\end{tabular}
\caption{Signal phase space restrictions for the T2K and MINERvA results.}
 \label{tab:phaseSpace}r
\end{table}
\end{center}

\section{Models}	
\label{sec:models}

The T2K and MINERvA results are compared to the predictions of models available in some of the latest neutrino event generators, including those which are (or are at least very similar to) those used as the inputs to T2K~\cite{Abe:2017vif} and NOvA~\cite{NOvA:2018gge} oscillation analyses: NEUT 5.4.0~\cite{Hayato:2002sd, Hayato:2009zz} and GENIE 2.12.4~\cite{Andreopoulos:2009rq}. All of these event generators describe the primary neutrino nucleus interaction with an inclusive cross-section model. Semi-inclusive predictions are then obtained by: selecting the motion of an initial state nucleon momentum from a predicted distribution of Fermi motion; conserving energy and momentum in the primary interaction; and then propagating the final state hadrons through an FSI model. Although this approach of factorising the components of the interaction is certainly not guaranteed to give similar results to a full semi-inclusive calculation~\cite{Moreno:2014kia}, it allows a tractable prediction of complicated final states. Moreover, full semi-inclusive predictions of neutrino-nucleus interactions are not yet available. 

NEUT 5.4.0 is capable of several different descriptions of CCQE interactions. It can simulate CCQE events according to: the Llewellyn-Smith formalism~\cite{LlewellynSmith:1971uhs} based on a relativistic Fermi gas (RFG) model of the Fermi motion; the spectral function (SF) from Ref.~\cite{Benhar:1994hw}; and a local Fermi gas (LFG) 1p1h model based on the work of Nieves et. al in Ref.~\cite{Nieves:2011yp} (where the latter is its nominal configuration). In all of these the axial mass used for CCQE processes ($M_A^{QE}$) is set to $\sim1.0$ GeV. For LFG and RFG the effect of Random Phase Approximation (RPA) corrections, as computed in Ref.~\cite{Nieves:2011yp}, are applied. RPA is not applied to SF. Resonant pion production (RES) is described by the Rein Sehgal model~\cite{Rein:1980wg} with the axial mass $M_A^{RES}$ set to 1.21 GeV whilst the simulation of 2p2h interactions is based on the model from Nieves et. al in Ref.~\cite{Nieves:2011yp}, both of which use an RFG model of the Fermi motion. Although the 2p2h contribution should be different in the SF description of CCQE with respect to what has been calculated in Ref.~\cite{Nieves:2011yp} for a Fermi gas, a more suitable computation is not yet available in simulations so the same 2p2h contribution from Nieves et. al. is added on top of SF. The deep inelastic scattering (DIS), relevant at high neutrino energy, is modelled using the parton distribution function GRV98~\cite{Gluck:1998xa} with corrections by Bodek and Yang~\cite{Bodek:2003wd}. The FSI, describing the transport of the hadrons produced in the elementary neutrino interaction through the nucleus, are simulated using a semi-classical intranuclear cascade model.


GENIE 2.12.4 also uses the Llewellyn-Smith model for CCQE events, but utilises a model of the Fermi motion based on an RFG with modifications from Bodek and Ritchie~\cite{Bodek:1981wr}. The GENIE `empirical MEC' model is used to describe 2p2h interactions. In the GENIE simulation used here, FSI is described by its nominal empirical model (known as `hA') which allows easy re-weighting of different FSI components. RES interactions are described using the model of Rein and Sehgal.

In addition to the NEUT and GENIE, the GiBUU 2017 theory framework is also compared to some results. GiBUU~\cite{Buss:2011mx} uses the Giessen-Boltzmann-Uehling-Uhlenbeck implementation of quantum-kinetic transport theory to describe FSI and treats bound nucleons within coordinate- and momentum-dependent potential using an LFG momentum distribution. The CCQE process is modelled as in Ref.~\cite{Leitner:2006ww} with $M_A^{QE}=1.03$~GeV. The 2p2h contribution is simulated by considering only the transverse contributions and translating the response measured in electron scattering to the neutrino case~\cite{Gallmeister:2016dnq}. The model used for single pion production has its vector part determined from an analysis of electro-pion production data and its axial part from a fit to hydrogen/deuterium neutrino data~\cite{Mosel:2017nzk}. The DIS is simulated with PYTHIA v6.4~\cite{Sjostrand:2006za}.

The comparison of these models to measurements is performed using NUISANCE~\cite{Stowell:2016jfr}.

\section{Comparisons to results}
\label{sec:modelComp}

\begin{figure}[htbp!]
\begin{center}
\begin{overpic}[width=0.99\linewidth]{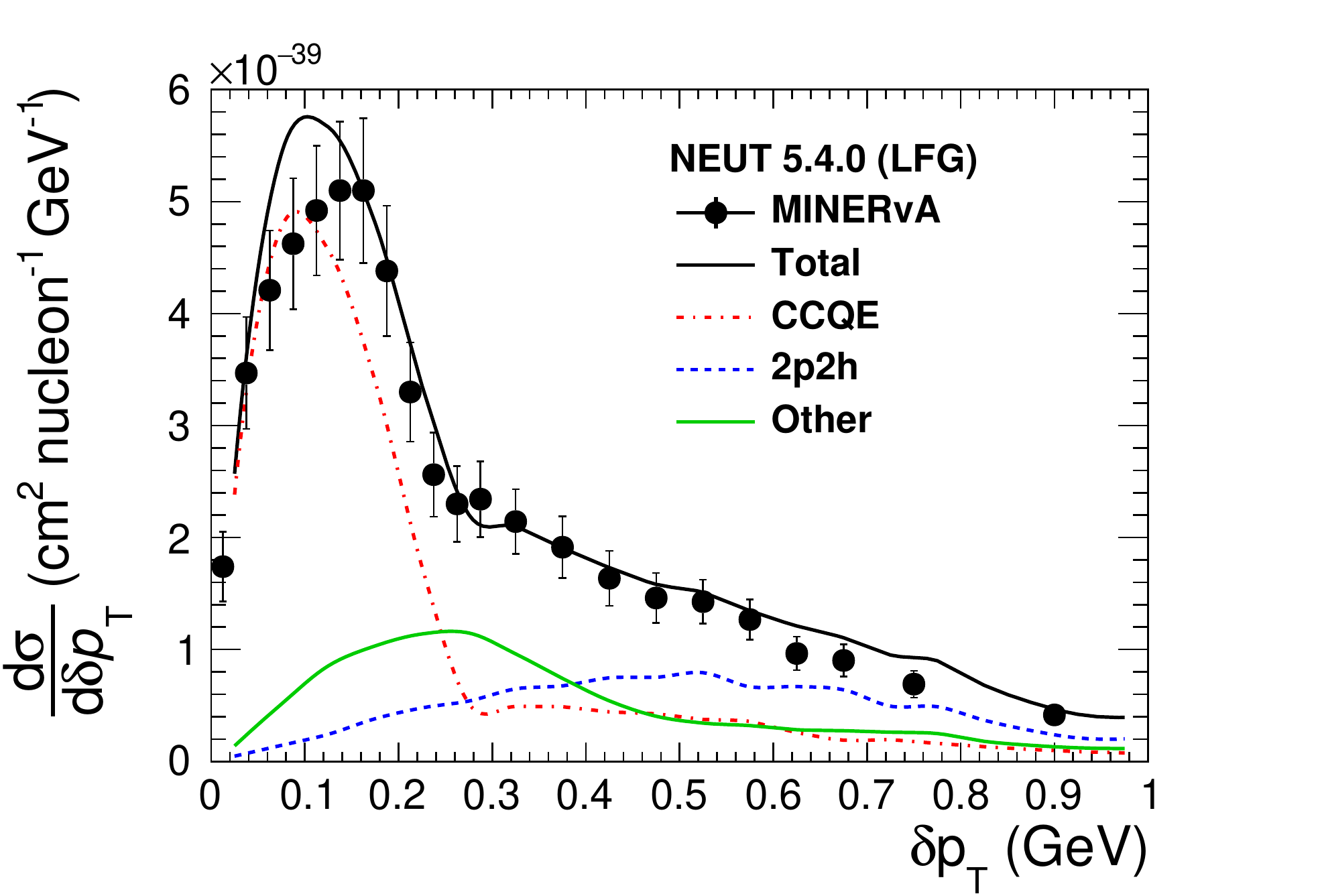}
   \put(52,30){$\chi^2_{NEUT}=62.1$}
\end{overpic}
\begin{overpic}[width=0.99\linewidth]{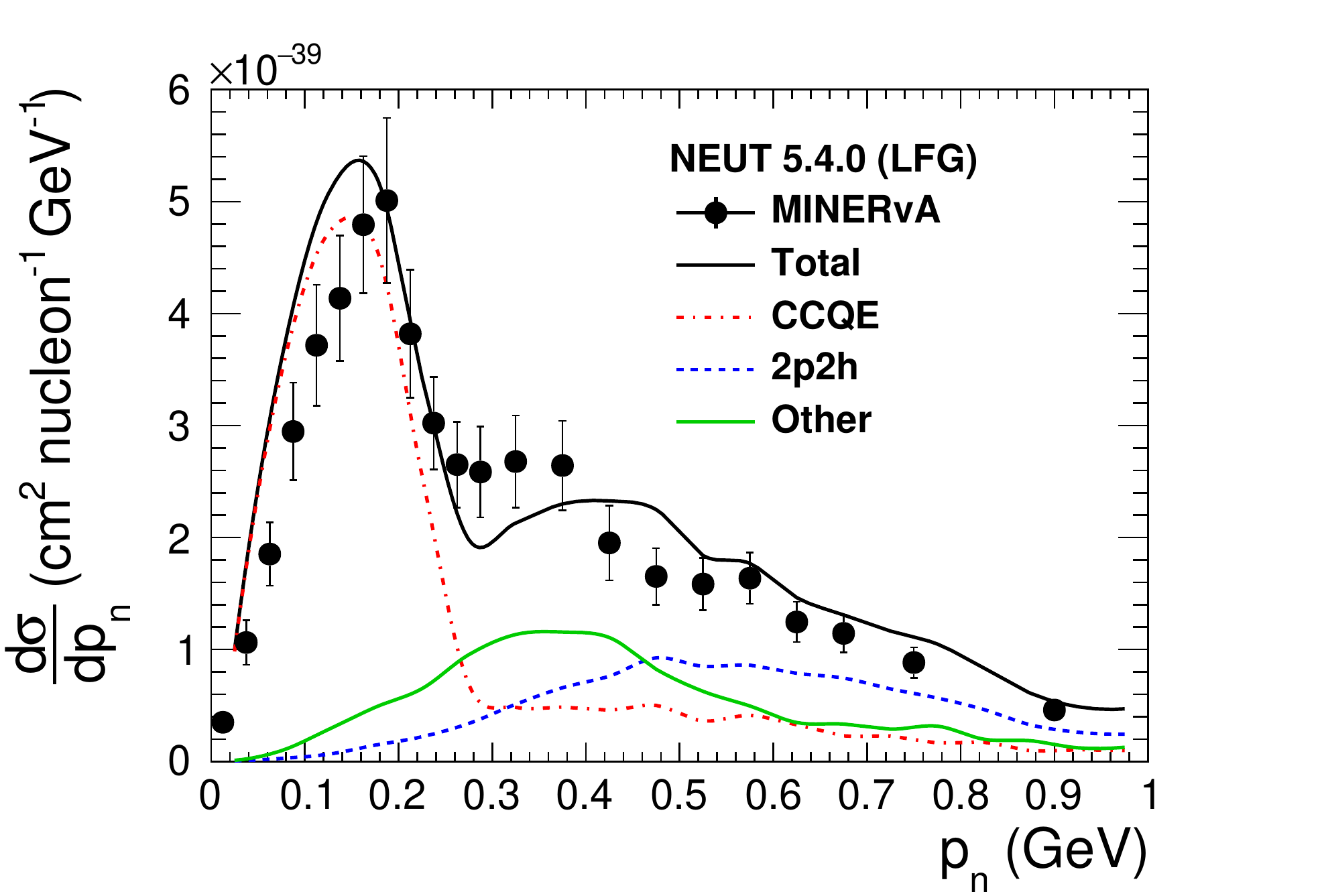}
   \put(52,30){$\chi^2_{NEUT}=122.2$}
\end{overpic}
\caption{MINERvA STV results in $\delta p_T$ and $p_n$ are shown alongside the NEUT 5.4.0 prediction (see Sec.~\ref{sec:models} for details) broken down by interaction model where the $\chi^2$ of the comparison is also shown.}
\label{fig:ModeComp_M}
\end{center}
\end{figure}

\begin{figure}[htbp!]
\begin{center}
\begin{overpic}[width=0.99\linewidth]{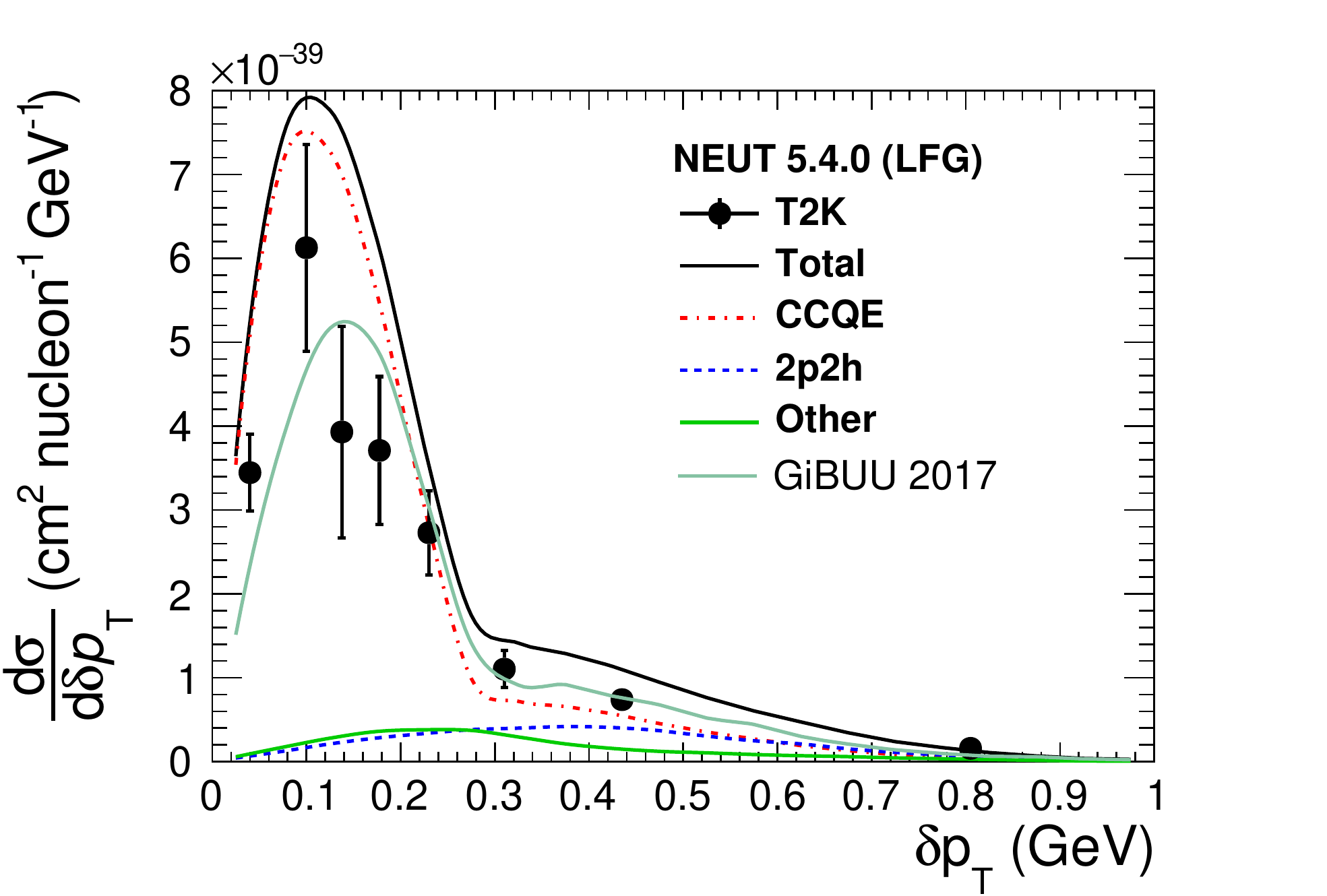}
   \put(52,25){$\chi^2_{NEUT}=31.4$}
\end{overpic}
\begin{overpic}[width=0.99\linewidth]{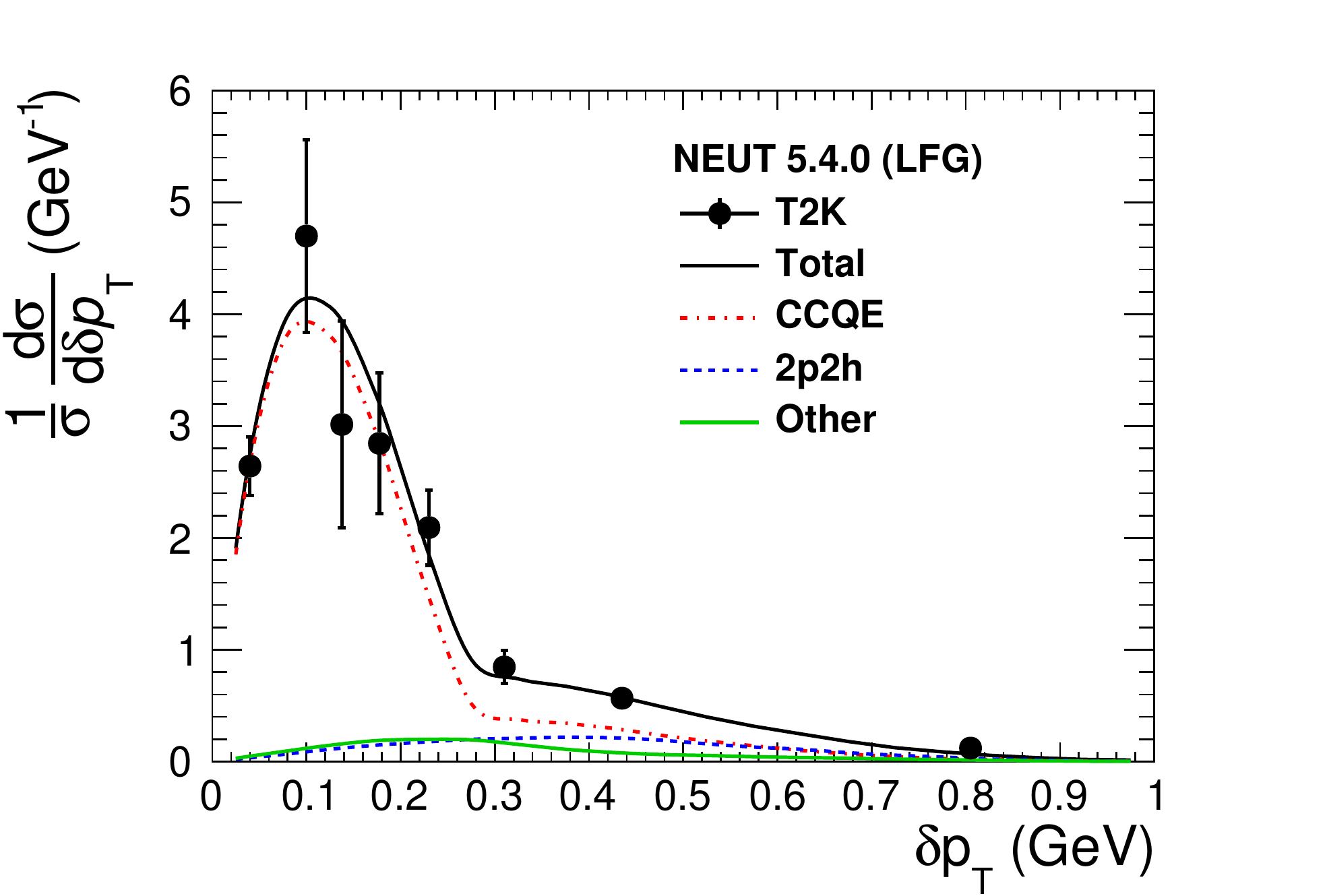}
   \put(52,30){$\chi^2_{NEUT}=3.3$}
\end{overpic}
\caption{The T2K full and shape-only $\delta p_T$ are shown alongside the NEUT 5.4.0 prediction (see Sec.~\ref{sec:models} for details) broken down by interaction model where the $\chi^2$ of the comparison is also shown. The GiBUU 2017 prediction is also shown as an alternative model for the full result.}\label{fig:ModeComp_T}
\end{center}
\end{figure}

Firstly the NEUT 5.4.0 nominal predictions (using an LFG model for CCQE interactions) in $\delta p_T$ and $p_n$ are compared to the MINERvA results in Fig.~\ref{fig:ModeComp_M} and to the $\delta p_T$ T2K results in Fig.~\ref{fig:ModeComp_T}, both for the full cross section (where the GiBUU 2017 prediction is also shown for comparison) and shape-only cases. In general it can be seen that measurements of both $\delta p_T$ and $p_n$ are able to separate the QE contribution in the bulk from a tail containing 2p2h and other (almost entirely pion absorption) interaction modes. Because of MINERvA's higher energy neutrino beam, there is a much larger relative contribution from RES interactions and subsequent pion absorption in the tail. 

\begin{figure}[htbp!]
\begin{center}
\begin{overpic}[width=0.99\linewidth]{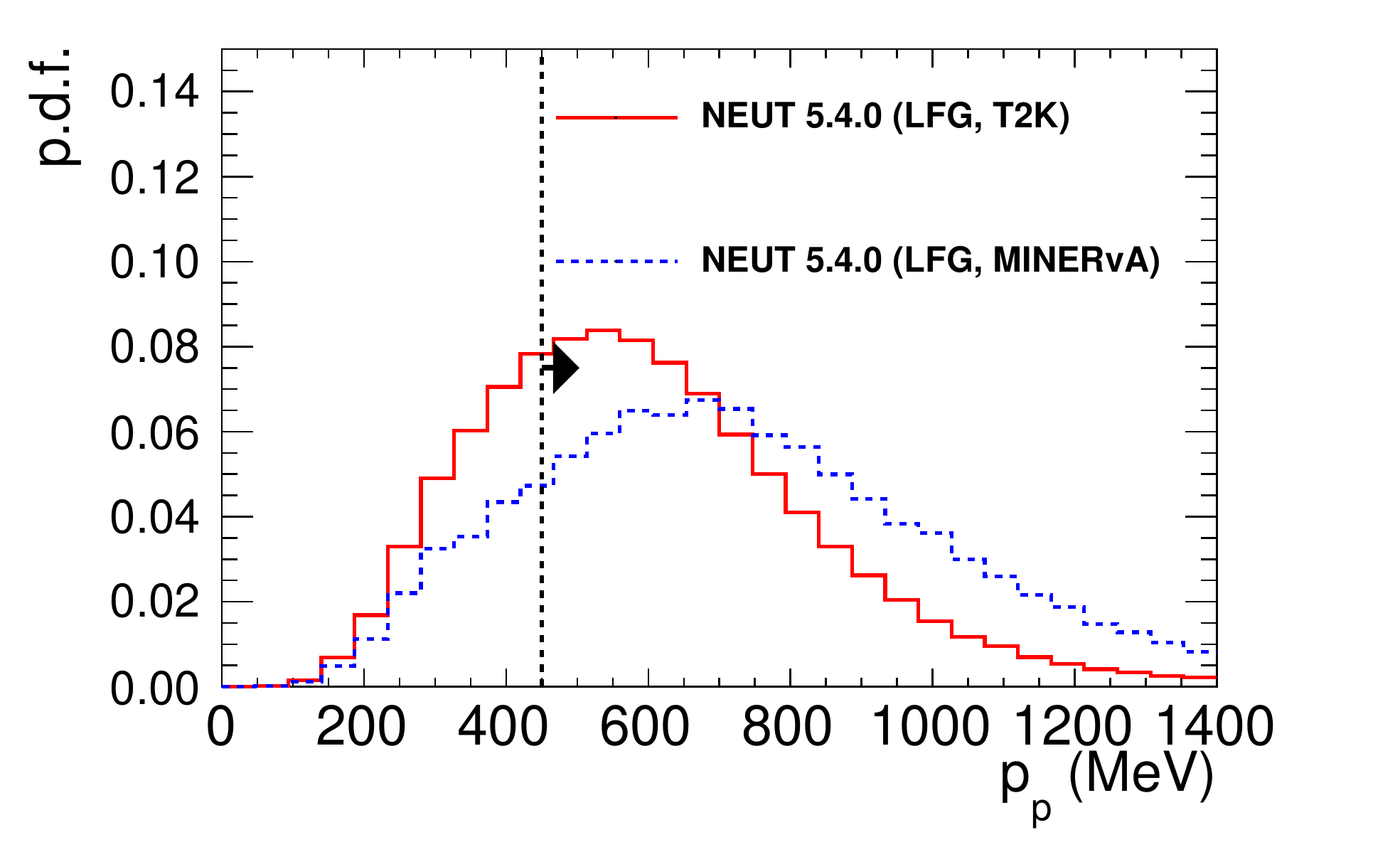}
\end{overpic}
\caption{The NEUT 5.4.0 prediction (see Sec.~\ref{sec:models} for details) of the shape of the proton momentum distribution within the other kinematic phase-space constraints (see Tab.~\ref{tab:phaseSpace}) for T2K and MINERvA are shown alongside the position of the low proton momentum constraint (which is the same for both experiments).}
\label{fig:ppComp}
\end{center}
\end{figure}

Whilst NEUT appears to predict the overall normalisation well for the MINERvA results, this is not the case for T2K. A possible explanation for this is shown in Fig.~\ref{fig:ppComp}, which demonstrates that the phase-space constraint on proton momentum in the T2K analysis is close to a fairly sharp peak, meaning small alterations to the shape of this distribution can lead to large changes in the normalisation of the STV results whilst causing only small shape differences. As demonstrated in~\cite{Abe:2018pwo}, altering FSI can cause such a shift, in particular it was shown that stronger FSI in NEUT can give a better agreement with the normalisation of the T2K results. It is also known that NEUT's model does not well predict proton-nucleus scattering data, particularly at the lower proton energies relevant to these results~\cite{Ma:2017wxw}, so such FSI alterations may be reasonable. Moreover, Fig.~\ref{fig:ModeComp_T} also shows that the GiBUU prediction, which differs in its use of a more sophisticated FSI model, is in much better agreement with the overall normalisation, thereby further suggesting FSI as the source of the discrepancy (GiBUU and NEUT are in much better agreement for total CCQE-like cross section without any constraint on outgoing proton kinematics). A more detailed discussion of FSI is presented in Sec.~\ref{sec:fsi}, suffice to say here that shape-only comparisons to the T2K results (as in Fig.~\ref{fig:ModeComp_T}) partially allow a factorisation of FSI effects from those that primarily drive shape variations in $\delta p_T$ and $\delta \phi_T$ (such as 2p2h and Fermi motion) and so will be shown throughout the comparisons to the results. 


Considering both the shape-only T2K and the full MINERvA comparisons, the NEUT prediction appears accurate in the 2p2h-enhanced tail region in both. The CCQE bulk of the distribution, which is controlled mostly by the shape of the initial-state nucleon momentum distribution~\cite{Lu:2015tcr, Furmanski:2016wqo}, is well described in the T2K case, but there appears to be an offset in the rising edge for the MINERvA result, which will be further discussed in Sec.~\ref{sec:fm}. The transition region between the bulk and the tail is also described well for both T2K and MINERvA results in $\delta p_T$, but seems to under-predict in $p_n$, where the NEUT suggests a dip region because of the large offsets in the CCQE and Other peak positions which is not seen in the MINERvA result. The shift in the Other peak between $\delta p_T$ and $p_n$ likely comes from the fact that the unseen pions produced in RES interactions tend to be fairly forward-going, and therefore contribute a substantial $\delta p_L$. The exact position of the dominant RES component of the peak relies on, among other effects, an accurate prediction of the energy-momentum transfer in RES interactions, which is known to be poorly described in some results~\cite{McGivern:2016bwh}, and pion absorption FSI.

The $\chi^2$ shown on all the figures throughout this section are calculated using the covariance matrix provided in the experimental data releases using all bins reported in the analysis (for the MINERvA results the very high $\delta p_T$, $p_n$ and $\delta \phi_T$ bins are not shown in the figures) and are not normalised by the number of degrees of freedom. For $\delta p_T$ the T2K covariance is characterised by moderate anti-correlations between adjacent bins, peaking in the centre of the bulk ($\sim25\%-35\%$), thereby offering substantial shape freedom, and becoming fairly flat in the tail (such that the T2K uncertainty here is reasonably well characterised by only the error bars shown on the figures). Conversely, the MINERvA covariance is dominated by extremely strong correlations between adjacent bins, peaking in the bulk ($\sim90\%-95\%$) but remaining strong in the tail ($\sim80\%-90\%$). These correlations make model-comparisons to the results very difficult to judge by-eye, thereby making the $\chi^2$ absolutely essential to interpret the comparisons. This property of strong correlations in the MINERvA results and mild anti-correlations for T2K is broadly true for the other observables considered. All results have a general correlation between distant bins mostly stemming from the flux-normalisation uncertainty (which contributes around a 10\% almost fully correlated error for both results). Here the T2K regularised results are shown, but the same $\chi^2$ are also calculated with the unregularised results in Appendix~\ref{app:unregComp} and are shown to be very similar. 


\subsection{Fermi motion}
\label{sec:fm}

Under the pure CCQE with no FSI assumptions used to form $p_n$ (as discussed in Sec.~\ref{sec:measurements}), the shape of the bulk region of both $p_n$ and $\delta p_T$ is controlled only by the initial state-nucleon momentum distribution~\cite{Lu:2015tcr, Furmanski:2016wqo}. Therefore a measurement in this region represents a fairly direct probe of the Fermi motion. However, since neutrino interaction cross sections will be different for initial state target nucleons moving toward or away from the incoming neutrino, and also because of the kinematic phase space constraints in each analysis (shown in Tab.~\ref{tab:phaseSpace}), the underlying nucleon momentum distribution is not sampled uniformly or even equivalently between T2K and MINERvA. For instance, in the T2K case seeing a proton above the 450~MeV/c tracking threshold from a peak incoming neutrino energy of 600~MeV is significantly more likely if the proton already has a significant initial-state momentum component along the direction of an interactions momentum transfer, whilst for MINERvA the higher energy beam and consequent larger typical momentum transfer to the proton (as shown by the higher-peaking MINERvA proton momentum distribution in Fig.~\ref{fig:ppComp}) means this effect is less significant.

\begin{figure}[htbp!]
\begin{center}
\begin{overpic}[width=0.99\linewidth]{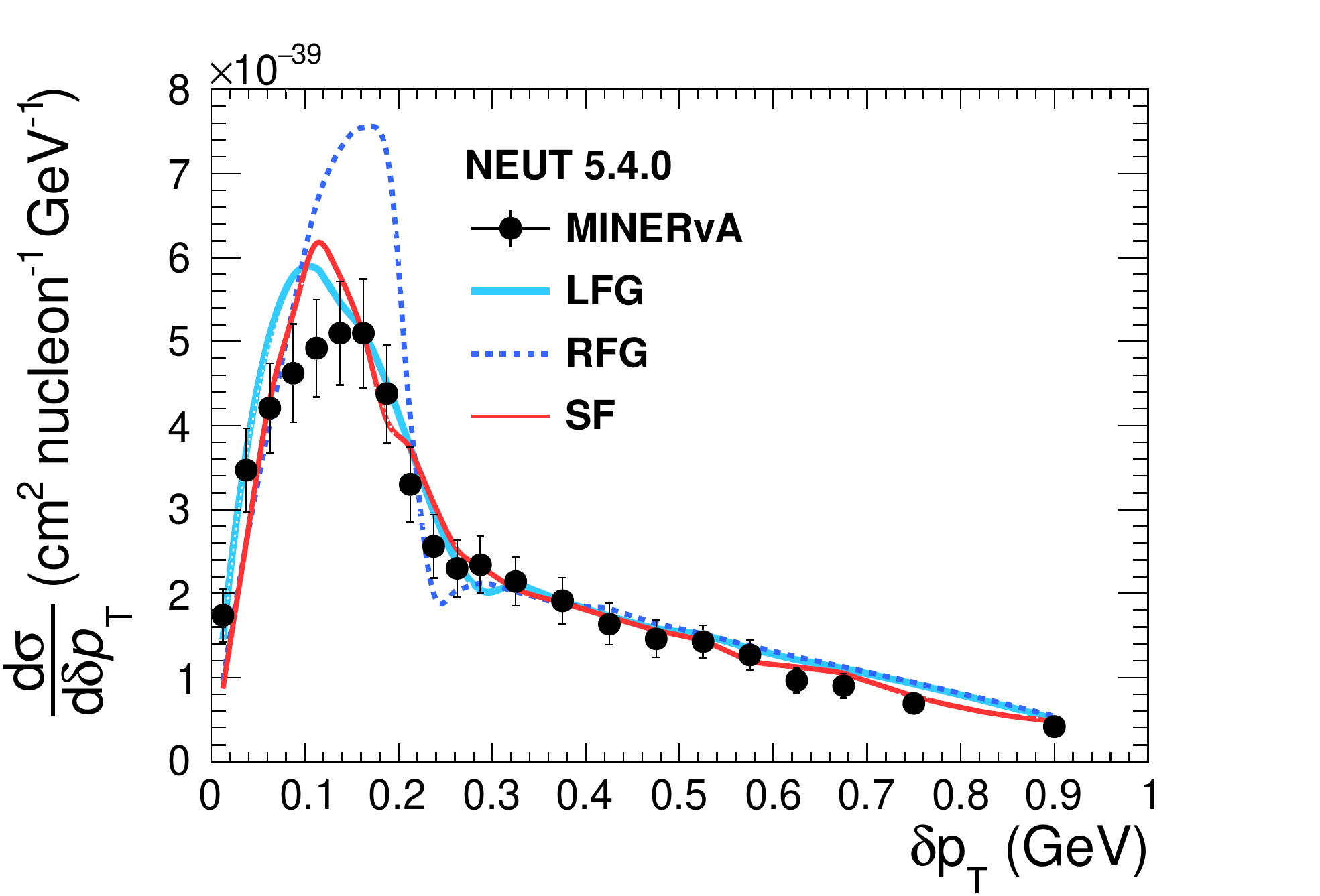}
   \put(54,45){$\chi^2_{LFG}=62.1$}
   \put(54,40.5){$\chi^2_{RFG}=321.5$}
   \put(54,36){$\chi^2_{SF}=104.9$}
\end{overpic}
\begin{overpic}[width=0.99\linewidth]{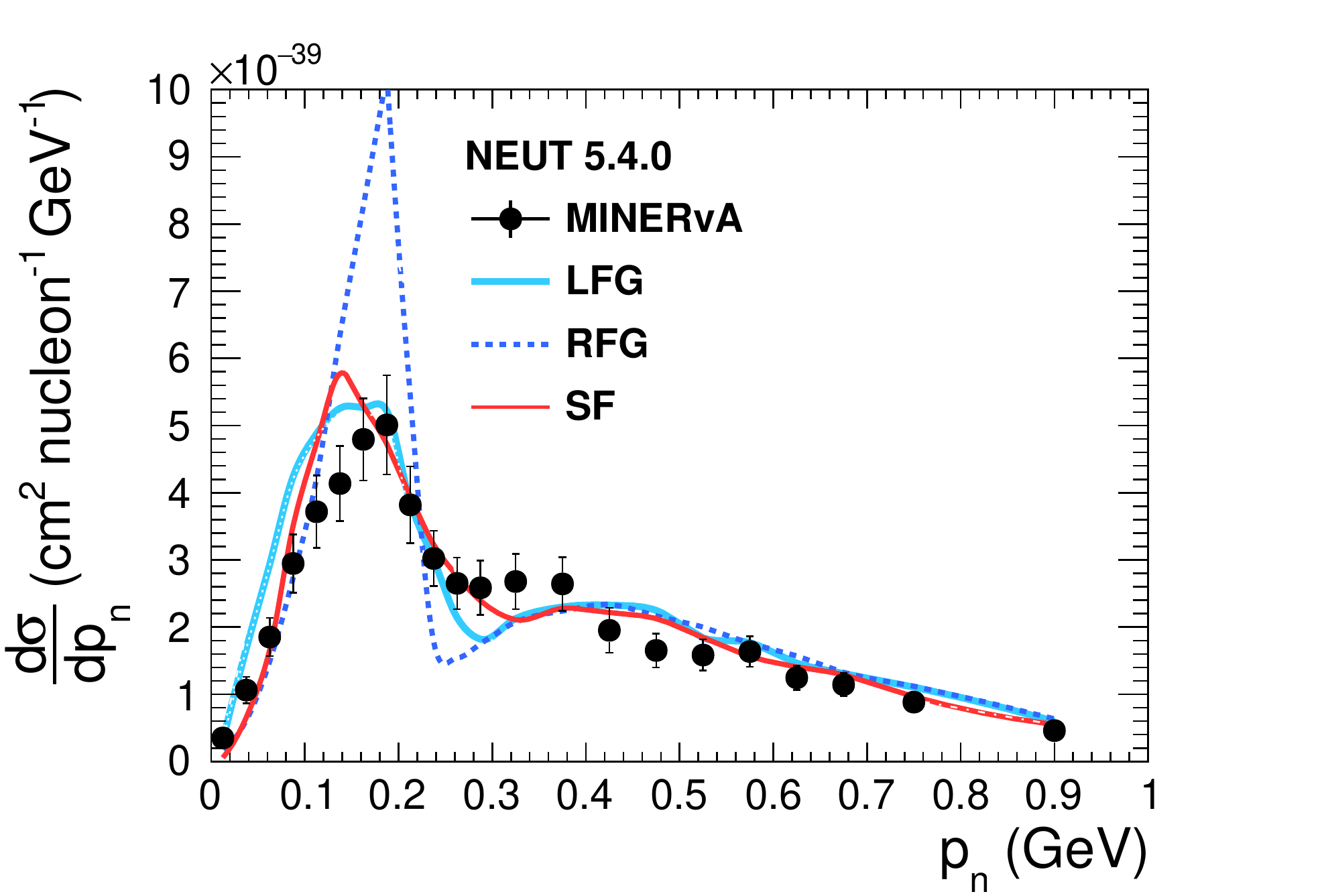}
   \put(54,46){$\chi^2_{LFG}=122.2$}
   \put(54,41.5){$\chi^2_{RFG}=309.1$}
   \put(54,37){$\chi^2_{SF}=110.8$}
 \end{overpic}
\caption{MINERvA STV results in $\delta p_T$ and $p_n$ are shown alongside the NEUT 5.4.0 prediction for different models of the Fermi motion (see Sec.~\ref{sec:models} for details). A $\chi^2$ of the comparison for each model is also shown.}
\label{fig:FMComp_M}
\end{center}
\end{figure}

\begin{figure}[htbp!]
\begin{center}
\begin{overpic}[width=0.99\linewidth]{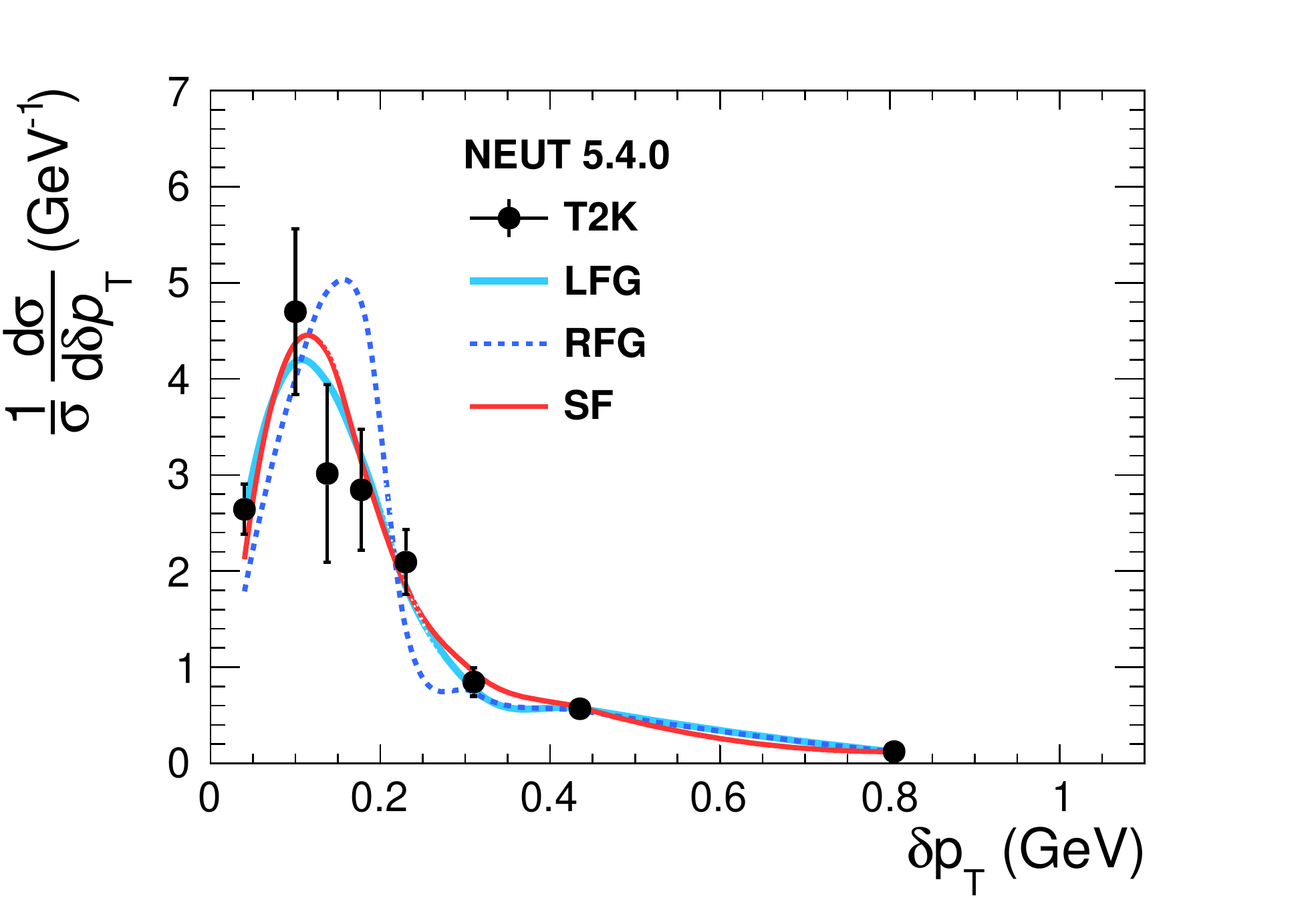}
   \put(54,46){$\chi^2_{LFG}=3.3$}
   \put(54,41.5){$\chi^2_{RFG}=45.8$}
   \put(54,37){$\chi^2_{SF}=10.0$}
 \end{overpic}
\caption{The T2K $\delta p_T$ shape-only result is shown alongside the NEUT 5.4.0 prediction for different models of the Fermi motion (see Sec.~\ref{sec:models} for details). A $\chi^2$ of the comparison for each model is also shown.}
\label{fig:FMComp_T}
\end{center}
\end{figure}

In Figs.~\ref{fig:FMComp_M} and~\ref{fig:FMComp_T} the three models of the Fermi motion within NEUT are compared to the MINERvA and T2K results respectively. A summary of the $\chi^2$ statistics calculated from these comparisons is given in Tab.~\ref{tab:chi2FM}. From these it can clearly be seen that the widely used RFG model is absolutely disfavoured by all the results. For the T2K case both LFG and SF describe the shape of the result well but there is a weak preference for the former. For the MINERvA analysis it can be seen that no model is able to provide a complete description of the result. In particular in both the $\delta p_T$ and $p_n$ comparisons demonstrate that both SF and LFG struggle to describe the rising edge of the bulk, whilst in $p_n$ only SF is able to describe the bulk-tail transition region, where the short-range correlations present in the model~\cite{Benhar:1994hw} give a larger tail to the Fermi motion, thereby filling in some of the aforementioned dip. 



\begin{center} 
\begin{table}[htbp!]
\begin{tabular}{ |l|c|c|c|c| } 
 \hline
  & $\chi^2_{LFG}$ & $\chi^2_{RFG}$ & $\chi^2_{SF}$ & $N_{bins}$  \\
 \hline
T2K ($\delta p_T$) & 31.4 & 129.6 & 21.8 & 8 \\
T2K SO ($\delta p_T$) & 3.3 & 45.8 & 10.0 & 8 \\
MINERvA ($\delta p_T$) & 62.1 & 321.5 & 104.9 & 24 \\
MINERvA ($p_n$) & 122.2 & 309.8 & 110.8 &  24\\
\hline
\end{tabular}
\caption{A summary of the $\chi^2$ calculated for comparisons of T2K and MINERvA STV results to NEUT 5.4.0 using different models for Fermi motion. The T2K $\chi^2$ shown here are calculated using the shape-only (SO) and full results. The number of bins in each result is also shown.}
 \label{tab:chi2FM}
\end{table}
\end{center}

\subsection{FSI and 2p2h}
\label{sec:fsi}

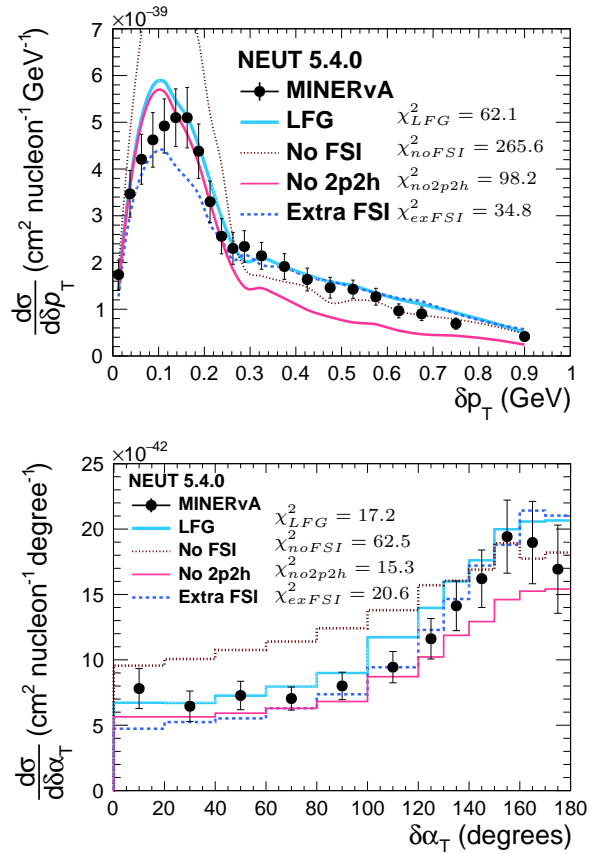
\begin{figure}[htbp!]
\vspace{-2.5mm}
\begin{center}
\begin{overpic}[width=0.99\linewidth]{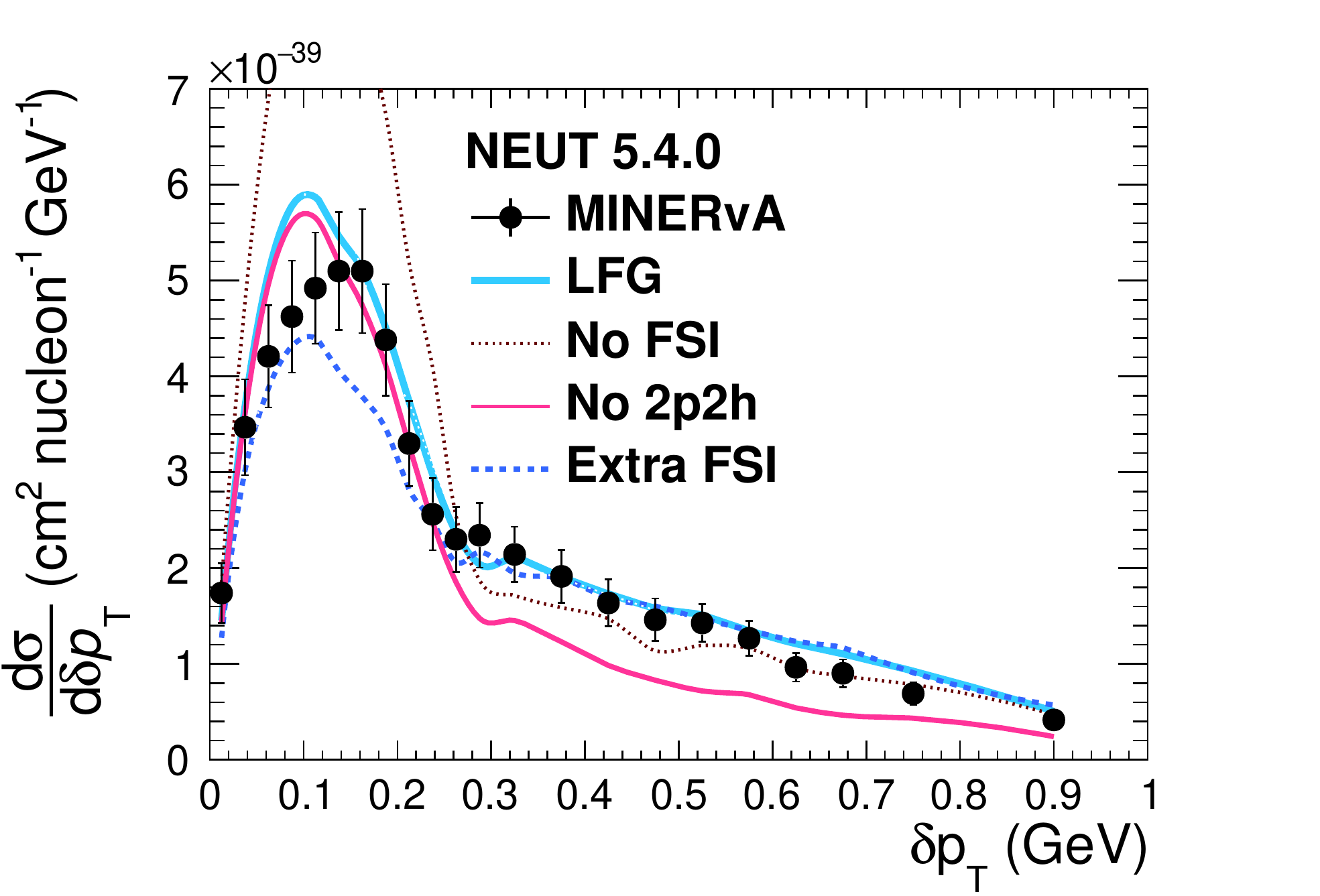}
   \put(60,47){\scriptsize$\chi^2_{LFG}=62.1$}
   \put(60,42){\scriptsize$\chi^2_{noFSI}=265.6$}
   \put(60,37){\scriptsize$\chi^2_{no2p2h}=98.2$}
   \put(60,32){\scriptsize$\chi^2_{exFSI}=34.8$}
\end{overpic}
\begin{overpic}[width=0.99\linewidth]{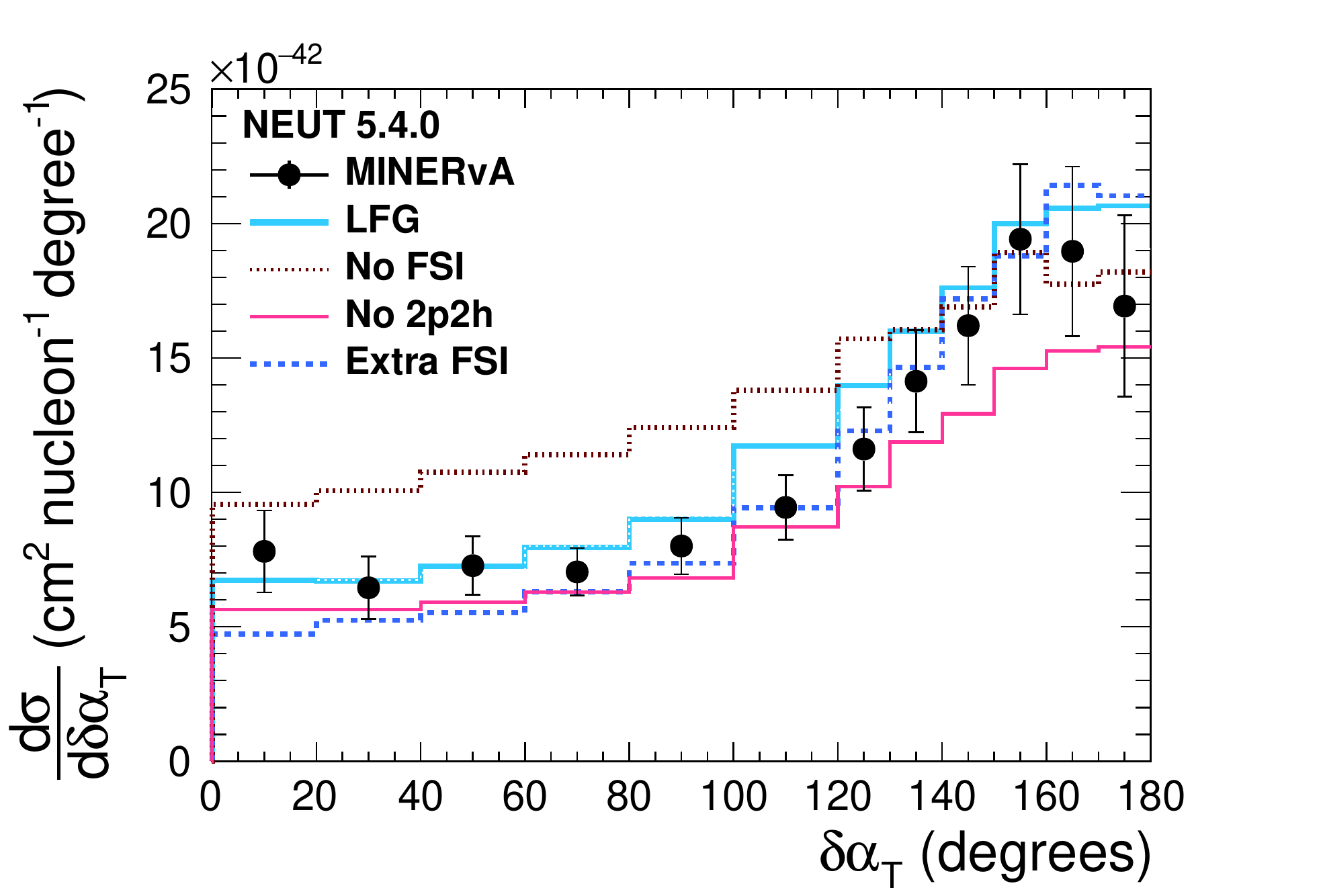}
   \put(41,52){\scriptsize$\chi^2_{LFG}=17.2$}
   \put(41,48){\scriptsize$\chi^2_{noFSI}=62.5$}
   \put(41,44){\scriptsize$\chi^2_{no2p2h}=15.3$}
   \put(41,40){\scriptsize$\chi^2_{exFSI}=20.6$}
 \end{overpic}
\caption{MINERvA STV results in $\delta p_T$ and $\delta \alpha_T$ are shown alongside the NEUT 5.4.0 prediction with and without 2p2h and with the nominal prediction (LFG, with 2p2h) following either the removal or strengthening (see the text for details) of nucleon FSI. The no FSI prediction in the full cross-section peaks at $\sim8.5\times10^{-39}$ cm$^2$ Nucleon$^{-1}$ GeV$^{-1}$. A $\chi^2$ of the comparison for each model is also shown.}
\label{fig:2p2hComp_M}
\end{center}
\end{figure}

\begin{figure}[htbp!]
\vspace{-2.5mm}
  \centering
  \begin{tikzpicture}
    \node(a){\includegraphics[width=0.99\linewidth]{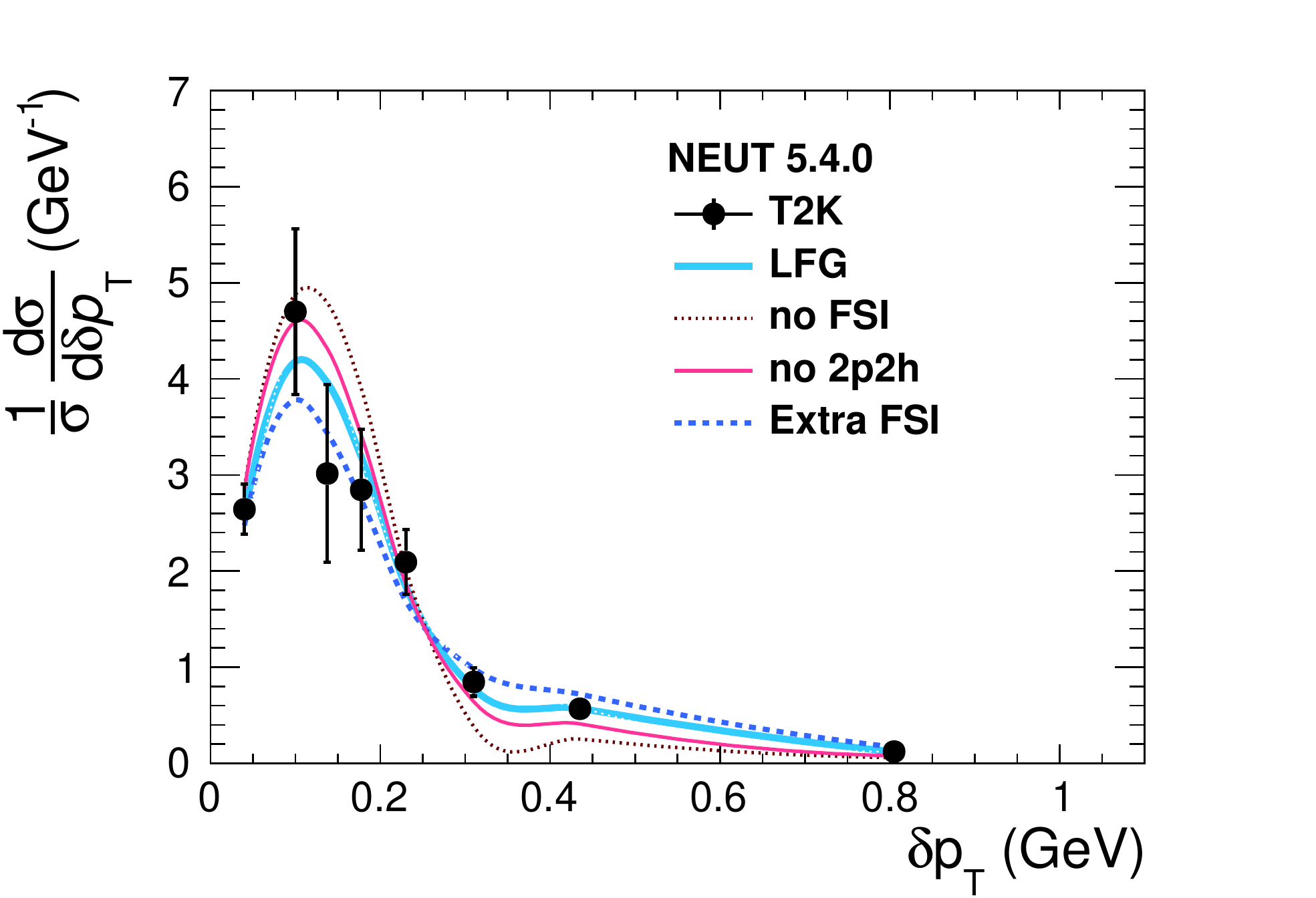}};
    \node at (a.south east)
    [
    anchor=center,
    xshift=-36.5mm,
    yshift=35mm
    ]
    {
        \includegraphics[width=0.335\textwidth]{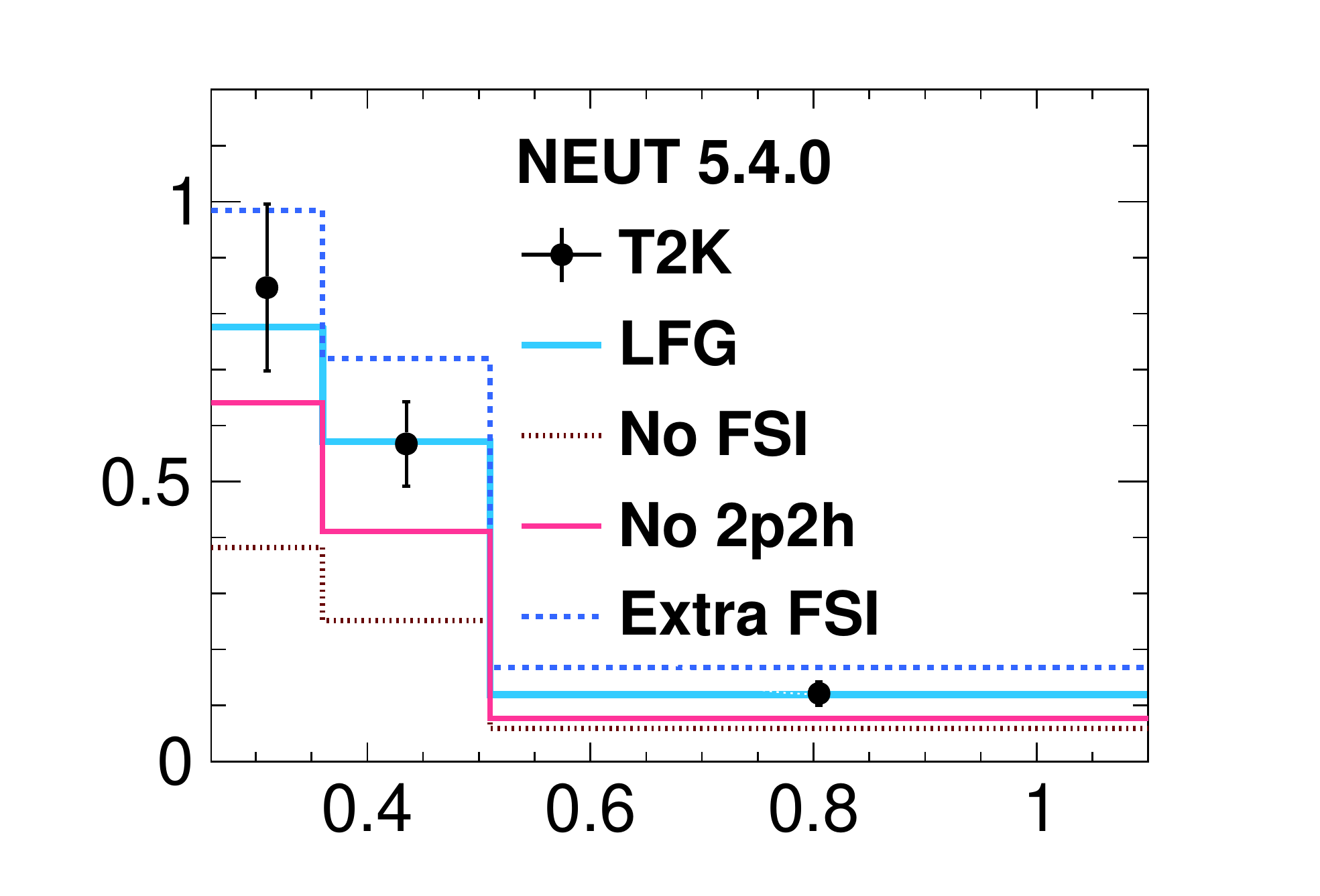}
    };
    \node[rotate=0] at (2.3,0.95) {\scriptsize {$\chi^2=3.3$}};
    \node[rotate=0] at (2.3,0.55) {\scriptsize {$\chi^2=88$}};
    \node[rotate=0] at (2.3,0.15) {\scriptsize {$\chi^2=29$}};
    \node[rotate=0] at (2.3,-0.25) {\scriptsize {$\chi^2=20$}};
    \end{tikzpicture}
    
    \vspace{-2.0mm}
    \begin{tikzpicture}
    \node(a){\includegraphics[width=0.99\linewidth]{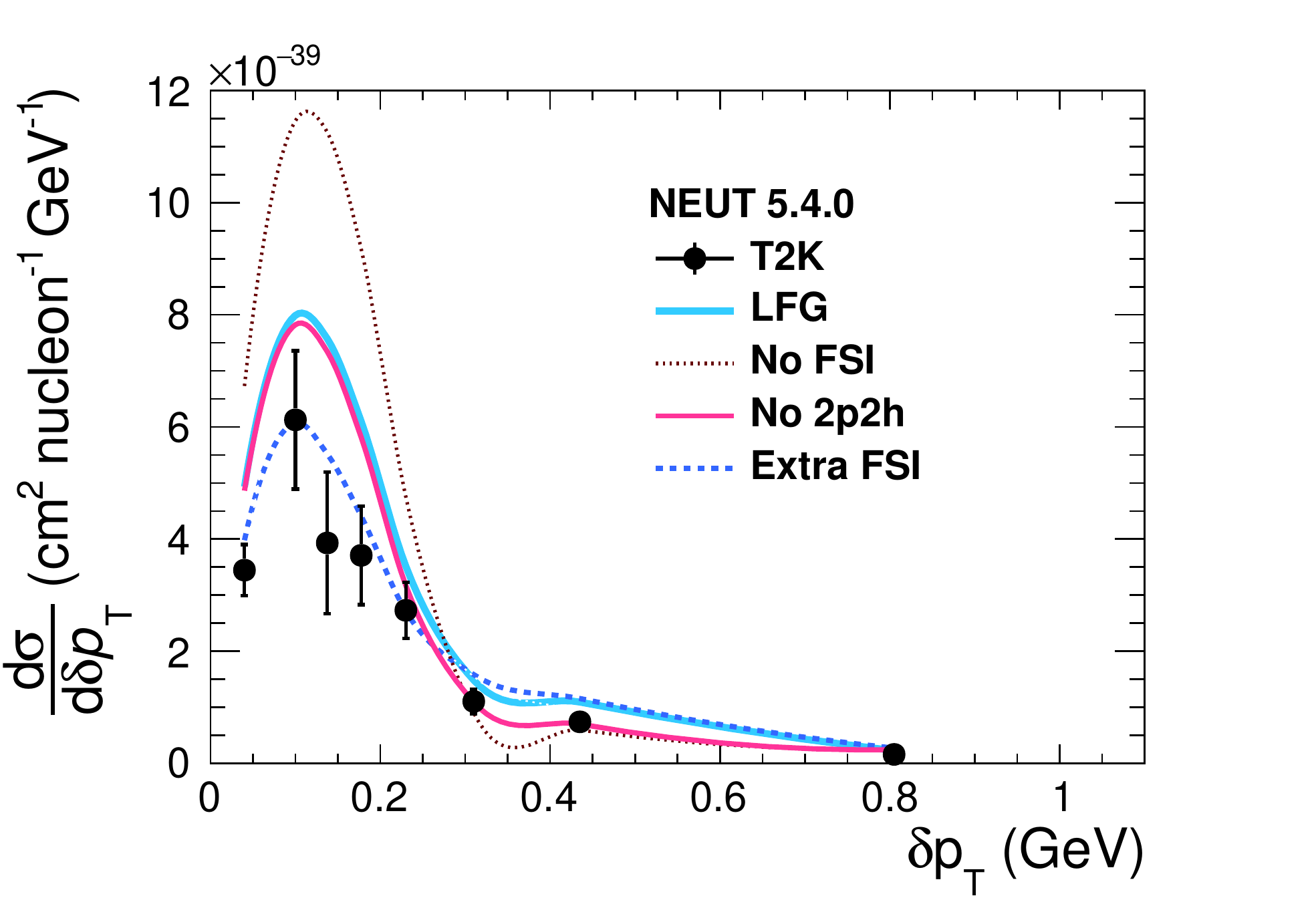}};
    \node at (a.south east)
    [
    anchor=center,
    xshift=-35.5mm,
    yshift=33.5mm
    ]
    {
        \includegraphics[width=0.318\textwidth]{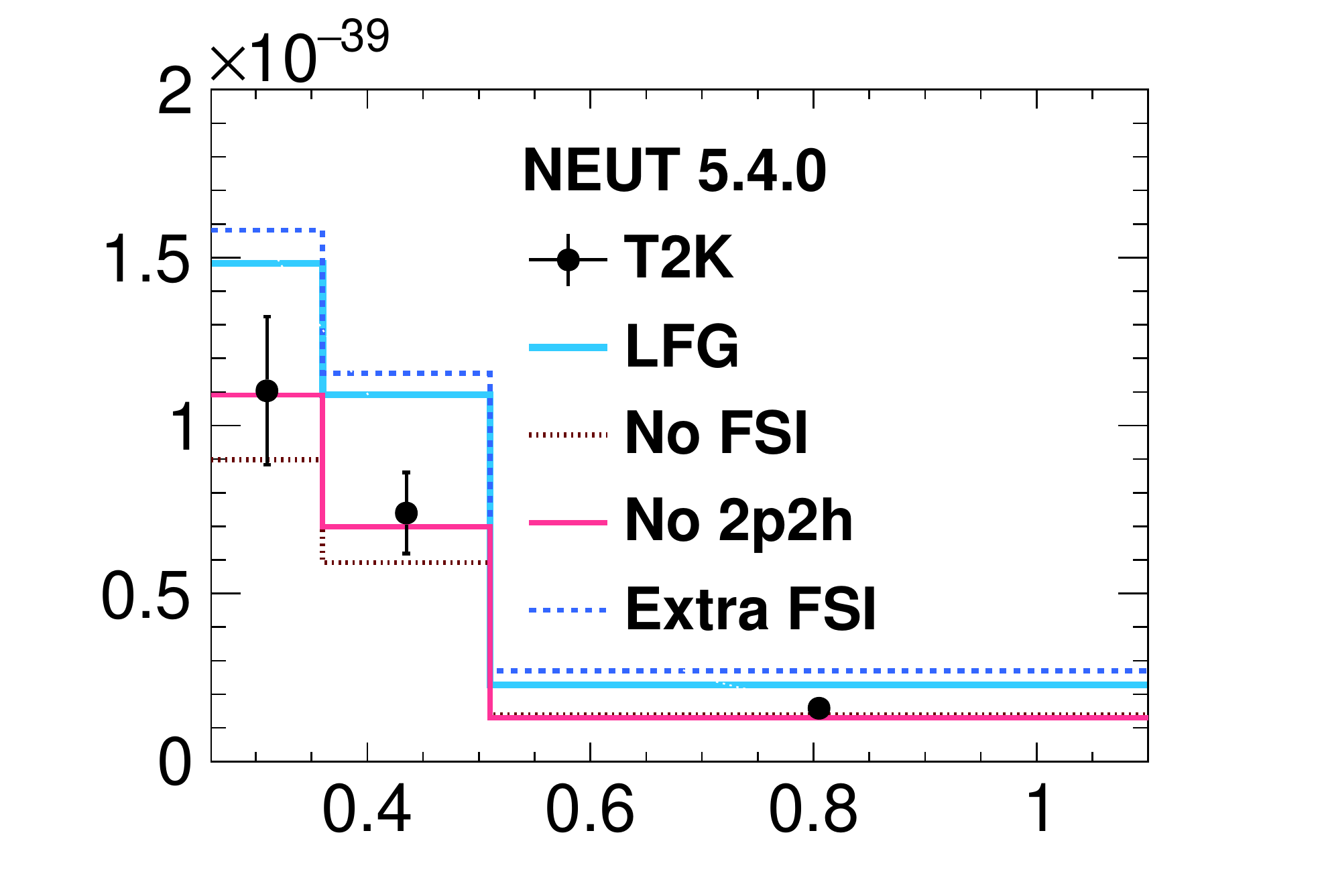}
    };
    \node[rotate=0] at (2.3,0.85) {\scriptsize {$\chi^2=31$}};
    \node[rotate=0] at (2.3,0.45) {\scriptsize {$\chi^2=372$}};
    \node[rotate=0] at (2.3,0.05) {\scriptsize {$\chi^2=62$}};
    \node[rotate=0] at (2.3,-0.35) {\scriptsize {$\chi^2=36$}};

    \end{tikzpicture}

\caption{The T2K full and shape-only results in $\delta p_T$ are shown alongside the NEUT 5.4.0 prediction with and without 2p2h and with the nominal prediction (LFG, with 2p2h) following either the removal or strengthening (see the text for details) of nucleon FSI. The inlays show close-ups of the tail. A $\chi^2$ of the comparison for each model is also shown. }
\label{fig:2p2hComp_T}
\end{figure}
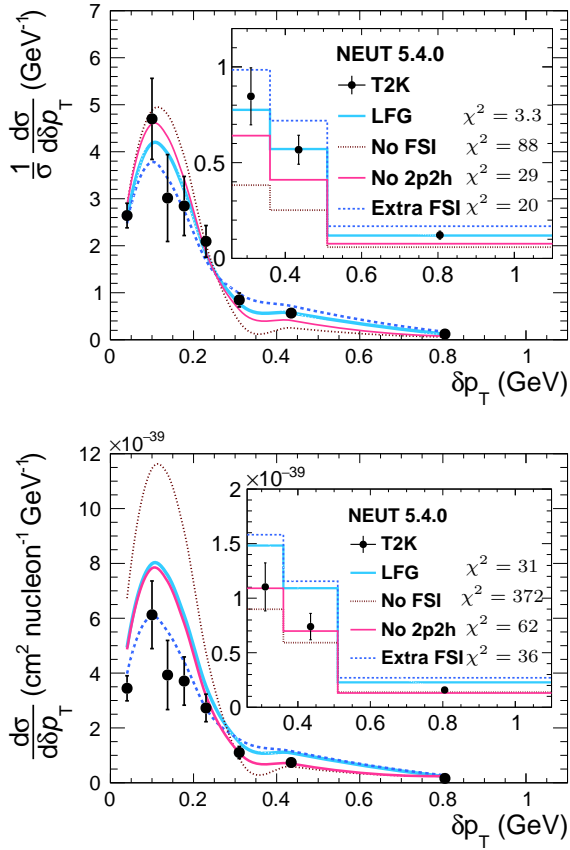

As already demonstrated in Figs.~\ref{fig:ModeComp_M} and~\ref{fig:ModeComp_T}, the tail of $\delta p_T$ is predicted to be strongly enhanced in 2p2h interactions and hence may be able to offer some characterisation of their contribution to the CCQE-like cross section (within the kinematic constraints listed in Tab.~\ref{tab:phaseSpace}). However, as discussed in~\cite{Lu:2015tcr}, the tail of $\delta p_T$ (and also $\delta \alpha_T$) may also be sensitive to FSI alterations, potentially in a way that is degenerate with variations of 2p2h. To asses the results sensitivity to these effects, and to evaluate whether 2p2h and FSI can be separated at all, Figs.~\ref{fig:2p2hComp_M}~and~\ref{fig:2p2hComp_T} show a comparison of the NEUT predictions for various modifications to 2p2h and FSI to the T2K and MINERvA measurements of $\delta p_T$ and (for MINERvA) $\delta \alpha_T$.  Although T2K also measures $\delta \alpha_T$ the exclusion of such a large proportion of low momentum protons due to the phase-space constraints (shown in Fig. ~\ref{fig:ppComp}) makes the result less sensitive~\cite{Abe:2018pwo}. A summary of the $\chi^2$ comparisons is shown in Tab.~\ref{tab:chi22p2h} where the $\chi^2$ calculated from the measurements of $p_n$, $\delta \alpha_T$ and $\delta \phi_T$ are also shown. Despite $\delta \phi_T$ being intrinsically less sensitive to nuclear effects than $\delta p_T$~\cite{Lu:2015tcr}, it does not require the measurement of momenta to reconstruct and therefore has substantially smaller experimental uncertainties. $\delta p_T$ is shown in the figures rather than $p_n$ as the former offers better statistical sensitivity to FSI and 2p2h (for these measurements), as can be seen in Tab.~\ref{tab:chi22p2h}. 


Fig.~\ref{fig:2p2hComp_T} shows that the NEUT nominal prediction (using an LFG model alongside 2p2h) is in excellent agreement with the T2K shape-only result, which also strongly disfavours the no 2p2h case. However, the normalisation discrepancy in the full result makes this difficult to interpret. A comparison to the full result shows that halving the mean free path between interactions in NEUT's nucleon FSI cascade (thereby effectively doubling FSI strength) brings the normalisation in the bulk of $\delta p_T$ into much better agreement, but in this case the tail becomes relatively stronger (due to stronger FSI also causing a larger proportion of events to have higher missing transverse momentum), thereby over-predicting the result and hence preferring a weakening of 2p2h. As previously discussed, comparisons of the NEUT FSI model to external nucleon scattering data suggest that there is scope for alterations, but it should be noted that this crude doubling of FSI strength represents an extreme variation and a more subtle approach would be required to really bring it into agreement with the external data. Nevertheless the overall comparison suggests that, within the kinematic phase-space considered by the analyses, 2p2h strength should not be much larger than that predicted by the Nieves et. al. model but may be weaker.

The comparisons to the MINERvA results in Fig.~\ref{fig:2p2hComp_M} show a more clear preference for 2p2h, demonstrating a need for its contribution in the $\delta p_T$ tail no matter whether the nucleon FSI is varied. It is also interesting to see that the stronger FSI that brings the bulk of $\delta p_T$ into agreement with the T2K result, is actually in much better agreement with the MINERvA result than the nominal NEUT prediction, supporting the previous suggestions for the requirement of some strengthening of NEUT's nucleon FSI. In particular the shape of the bulk is altered such that the rising edge is well predicted by the model, suggesting that an LFG model may be able to well describe the result with an altered FSI model. It can also be seen that the effect of 2p2h may be partially separated from FSI through a measurement of $\delta \alpha_T$, which shows a clear sensitivity to the latter but not the former in its shape (FSI is responsible for the vast majority of the large rise at high $\delta \alpha_T$, which is steeper for stronger FSI). However, whilst the experimental uncertainties on the current measurement offer a strong rejection of the no FSI prediction, there is only modest separation between the nominal and enhanced nucleon FSI models. 

It should also be noted that in NEUT's FSI model almost all CCQE-like interactions leave a proton in the final state, whilst in the GiBUU transport model around 7\% do not. Adding such a component into the NEUT model may help reduce the predictions normalisation to bring it into better agreement with the results. 

In addition to varying nucleon FSI, altering the possibility for pion absorption FSI to make RES events appear CCQE-like may also be partially degenerate with alterations to 2p2h. However, since NEUT's model is in fairly good agreement with external pion-scattering data across a wide range of kinematics~\cite{Ma:2017wxw}, which suggests only a small increase in pion absorption (by no more than around 25\%) may be plausible, there is less scope for modification. For the T2K comparisons such variations are expected to cause only a small increase in the `other' contribution to the tail in Fig.~\ref{fig:ModeComp_T}, as the lower energy neutrino beam produces only a few such interactions (a prediction shared by NEUT and GiBUU~\cite{Dolan:2018sbb}). For the MINERvA comparisons the shape of the pion-absorption contribution to $\delta p_T$ is predicted to be quite distinct from the 2p2h, contributing dominantly to the bulk-tail transition region (as seen in Fig.~\ref{fig:ModeComp_M}) where there is good agreement (although it should be remembered that the apparent shift in the $p_n$ peak suggests that the peak position may not be well modelled). In any case, the net effect of a small increase in pion absorption FSI would be to cause a similarly small increase in the prediction in the $\delta p_T$ tail, which would suggest the need for a slight weakening of 2p2h.

Overall comparisons of the nominal NEUT prediction to full MINERvA and shape-only T2K results suggest the Nieves et. al. model provides approximately the correct 2p2h strength. When the full T2K results are also considered, there is a consensus between the T2K and MINERvA results that is scope for some strengthening of FSI, which would then require a corresponding weakening on 2p2h to keep the prediction in line with the $\delta p_T$ tails of the T2K result. A more quantitative conclusion would require the provision of detailed uncertainties on NEUT's FSI model (for both pions and nucleons). These conclusions also rely on the NEUT prediction for how much of the 2p2h in the Nieves et. al. model falls within the proton kinematic phase space listed in Tab.~\ref{tab:phaseSpace} which, as discussed in Sec.~\ref{sec:models}, depends on ad-hoc additions to the original inclusive model. Avoiding this would require a semi-inclusive 2p2h model.


Further experimental results may also help: improving the precision of measurements of $\delta \alpha_T$ may facilitate the validation relevant FSI modelling which would then allow the primary unknown in the tail of $\delta p_T$ to be 2p2h. Furthermore, it may be interesting to also make a multi-differential cross-section measurement of $\delta p_T$ and $\delta \alpha_T$ simultaneously. In the rise of $\delta \alpha_T$ the tail of $\delta p_T$ would be dominated by FSI, whilst at small $\delta \alpha_T$ it would stem mostly from 2p2h, thereby potentially allowing an acute separation of the two. Such a measurement would require high statistics and a strong sensitivity to FSI through $\delta \alpha_T$ and may therefore be achievable by MINERvA or by the planned upgrade to the T2K near detector (which will have a much lower tracking threshold for protons)~\cite{Blondel:2299599}.

\begin{center} 
\begin{table}[htbp!]
\begin{tabular}{ |l|c|c|c|c|c| } 
 \hline
  & $\chi^2_{LFG}$ & $\chi^2_{no2p2h}$ & $\chi^2_{noFSI}$ & $\chi^2_{exFSI}$ & $N_{bins}$  \\
 \hline
T2K ($\delta p_T$) &  31.4 & 62.1 & 371.9 & 36.3 & 8 \\
MINERvA ($\delta p_T$) &  62.1 & 98.3 & 265.6 & 34.8 & 24 \\
T2K ($\delta \alpha_T$) &  60.5 & 34.9 & 107.5 & 65.3 & 8 \\
MINERvA ($\delta \alpha_T$) &  17.2 & 15.3 & 60.6 & 20.6 & 12 \\
T2K ($\delta \phi_T$) &  36.9 & 48.8 & 303.5 & 64.2 & 8 \\
MINERvA ($\delta \phi_T$) & 102.2 & 74.6 & 237.5 & 113.6 & 23 \\
MINERvA ($p_n$) &  122.2 & 137.7 & 293.5 & 106.2 & 24 \\
\hline
\end{tabular}
\caption{A summary of the $\chi^2$ calculated for comparisons of T2K and MINERvA STV results to NEUT 5.4.0 prediction with and without 2p2h and with the nominal prediction (LFG, with 2p2h) following either the removal or strengthening (see the text for details) of nucleon FSI. The number of bins in each result is also shown.}
 \label{tab:chi22p2h}
\end{table}
\end{center}

\subsection{FSI in GENIE}
\label{sec:geniefsi}

In Sec.~\ref{sec:fsi} a comparison of NEUT predictions to MINERvA and T2K STV results has suggested that the strength of the Nieves et. al. 2p2h model is approximately correct or too strong. This is in contrast to what has been observed by the MINERvA collaboration, which find that a significant empirical enhancement of the same Nieves et. al. 2p2h is required within the GENIE event generator in order to describe inclusive scattering data~\cite{Rodrigues:2015hik}. This may simply be because the difference between the true 2p2h strength and the model is quite different in the different kinematic phase spaces accessed by the STV measurements and the inclusive data (especially due to the aforementioned ad-hoc predictions required to evaluate the model's 2p2h strength in a specific region of proton kinematic phase space). However, the inclusive data reconstructs the total energy deposited in the MINERvA detector and so in addition to 2p2h it is also sensitive to both Fermi Motion (for which an RFG-based model is used in the MINERvA analysis) and the proportion of the energy which is carried away by (largely) invisible neutrons, which in turn is sensitive to FSI. Therefore it may be the apparent 2p2h enhancement is partially covering the potential lack of predictive power in the GENIE models used for FSI and Fermi motion.  

\begin{figure}[tbp!]
\begin{center}
\begin{overpic}[width=0.99\linewidth]{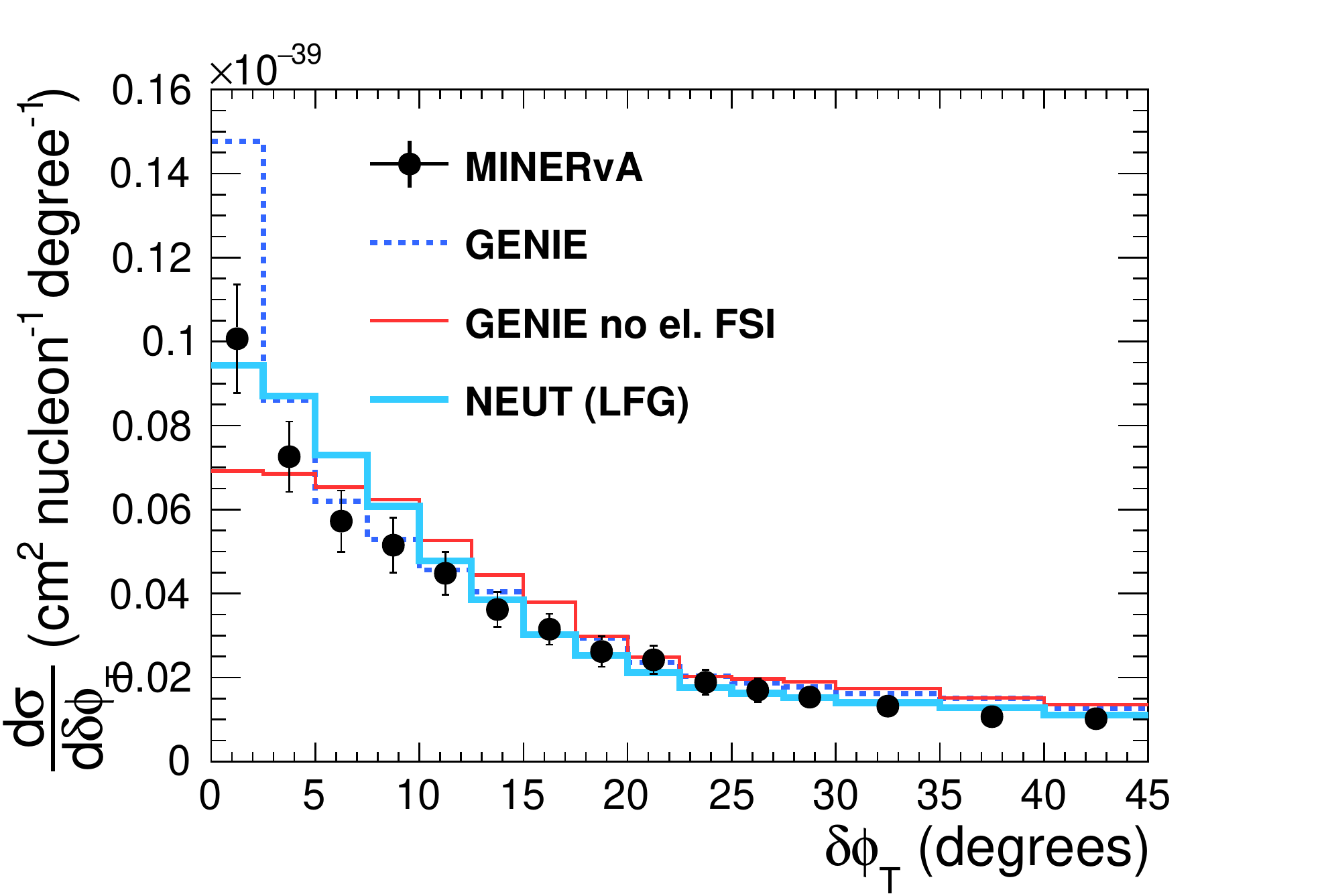}
   \put(55,48){$\chi^2_{GENIE}=213.4$}
   \put(60,42){$\chi^2_{noEl}=241.8$}
   \put(55,36){$\chi^2_{NEUT}=102.2$}
\end{overpic}
\begin{overpic}[width=0.99\linewidth]{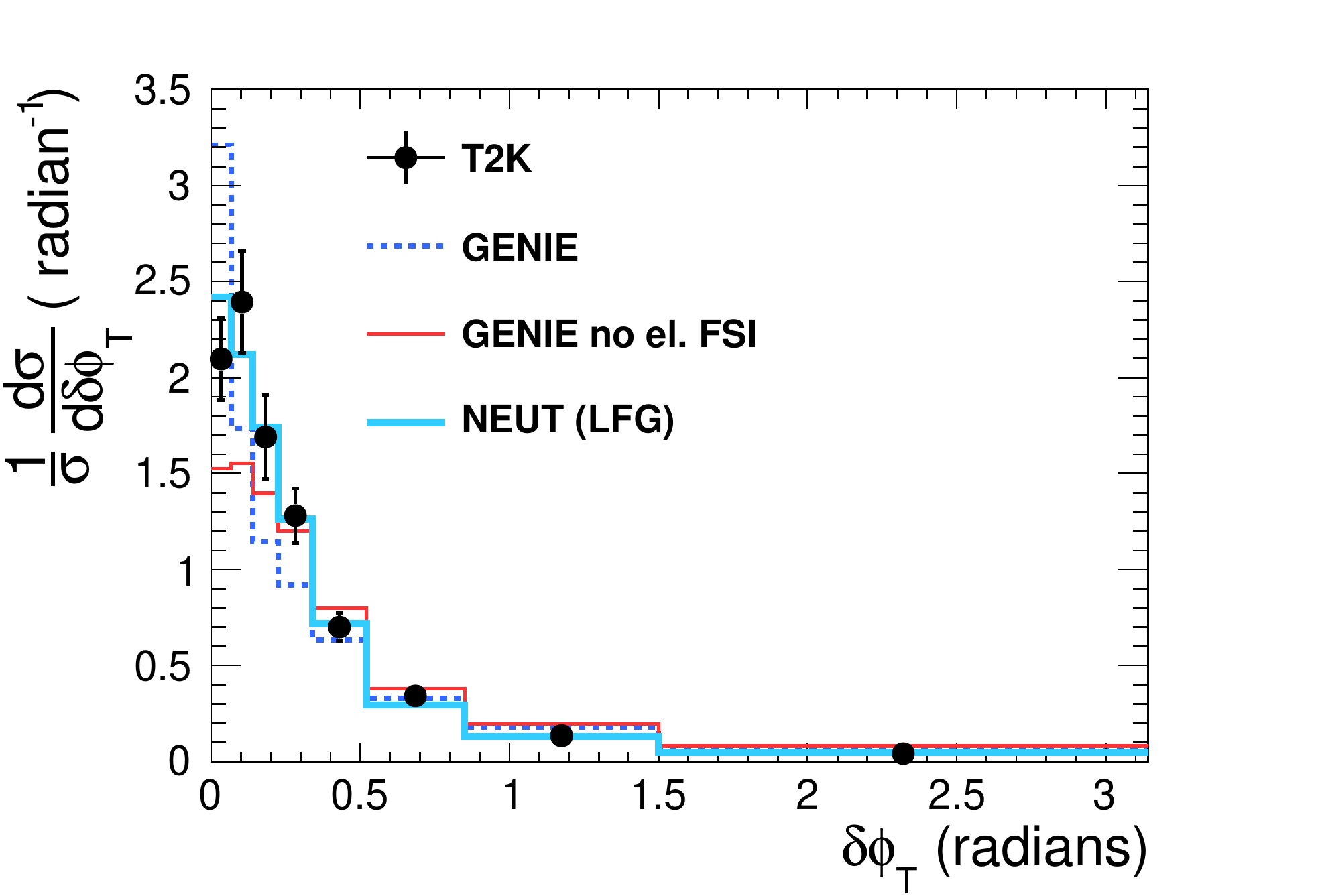}
   \put(55,49){$\chi^2_{GENIE}=122.8$}
   \put(60,42){$\chi^2_{noEl}=124.0$}
   \put(55,35){$\chi^2_{NEUT}=6.9$}
 \end{overpic}
\caption{T2K (shape-only) and MINERvA measurements of $\delta \phi_T$ are shown alongside the NEUT 5.4.0 prediction as well as that of GENIE 2.12.8 with and without elastic nucleon FSI. A $\chi^2$ of the comparison for each model is also shown. }
\label{fig:genieFSI}
\end{center}
\end{figure}
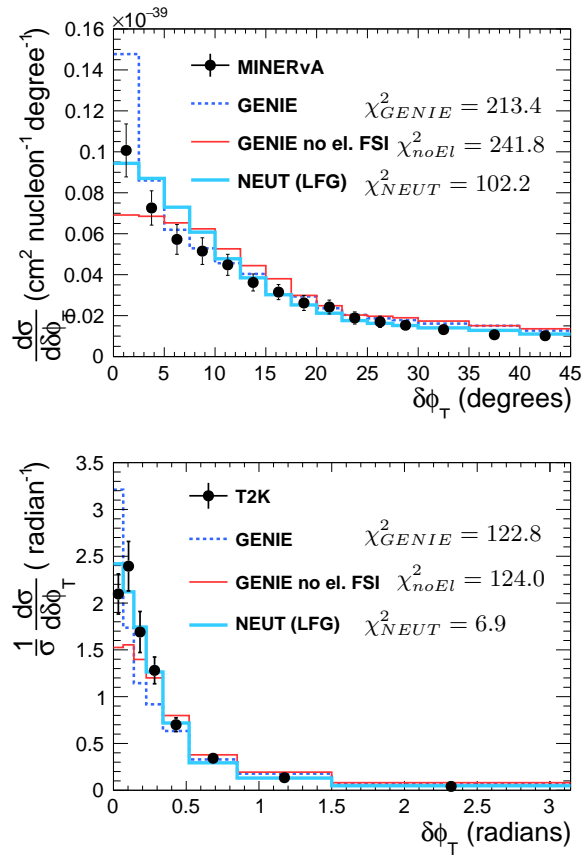

Here the GENIE's nominal `hA' FSI model, which is very different than NEUT's (as discussed in Sec.~\ref{sec:models}), is shown to be in stark disagreement with the MINERvA and T2K STV results, particularly emphasised by the measurements of $\delta \phi_T$, which are shown in Fig.~\ref{fig:genieFSI}. This disagreement is shown to stem from the elastic nucleon FSI in the GENIE empirical model predicting a very different distribution from those of other generators. This `anomaly' was first reported in~\cite{Lu:2015tcr} and the model has previously been compared to the T2K $\delta \phi_T$ results in~\cite{Abe:2018pwo}. The comparisons here also show that `turning off' the elastic FSI is still insufficient to describe the results and that a more sophisticated cascade or transport model (as is found in NEUT and GiBUU respectively) is likely necessary. In light of this it would be interesting to see whether a large empirical 2p2h enhancement would still be required to reproduce the MINERvA inclusive data if a more realistic base models of FSI and Fermi motion were to be used.

\section{Conclusion}

A simultaneous analysis of T2K and MINERvA measurements of kinematic imbalances between the outgoing muon and proton in CCQE-like interactions has allowed an in-depth exploration of the nuclear effects responsible for some of the largest systematics in neutrino oscillation measurements. The widely used RFG model of the Fermi motion has been firmly disfavoured by both results and, for NEUT's nominal FSI predictions, there is a clear joint preference for the presence of 2p2h final states with a strength consistent with the predictions of the Nieves et. al. model. However, this conclusion is somewhat degenerate with alterations to NEUT's FSI model and also relies on the prediction of proton kinematics from 2p2h interactions, which are generated using ad-hoc additions to the model. It has also been demonstrated that both measurements strongly disfavour the empirical `hA' FSI model used by the widely-used GENIE simulation, suggesting the need for a more sophisticated cascade or transport model. Overall these results show clear sensitivity to variations of key nuclear effects, but drawing more comprehensive conclusions requires the development of models (complete with uncertainties) capable of making semi-inclusive predictions with greater predictive power.

\vspace{-1.5mm}
\section*{Acknowledgements}
\vspace{-1.5mm}
\noindent I would like to thank S.~Bolognesi, C.~Wret, C.~Wilkinson, L.~Pickering, L.~Munteanu, K.~McFarland, K.~Mahn, G.~Megias, M.~Buizza-Avanzini and C.~Riccio for helpful discussions related to this manuscript. I am very grateful to the T2K and MINERvA Collaborations for producing and analysing the results used here, in particular to X.~Lu for patiently explaining many of the details of the MINERvA result. I also thank E.~Game for spotting several English errors in my initial draft. I acknowledge the support of CEA, CNRS/IN2P3 and P2IO in France and I am grateful to the MSCA-RISE project JENNIFER, funded by EU grant n.644294, for supporting EU-Japan mobility.

\appendix
\section{Comparison with unregularised results}
\label{app:unregComp}

T2K measured both unregularised and regularised differential cross sections in the STV. The unregularised results contain larger anti-correlations between adjacent bins (making the resut difficult to interpret by-eye) but should be less biased to the input simulation used by T2K. In the main body of this manuscript all comparisons have been shown with the regularised results, but in Tab.~\ref{tab:chi2regcomp} $\chi^2$ comparisons are also made to the unregularised shape-only and full results. These numbers are always similar, indicating that the physics conclusions here are not dependent on the use of regularisation in T2K's analysis. MINERvA have only released regularised results.


\makeatletter\onecolumngrid@push\makeatother

\begin{table}[t]
\centering
\begin{tabular}{ |l|c|c|c|c|c|c|c|c| } 
 \hline
  & $\chi^2_{LFG}$ &$\chi^2_{no2p2h}$ & $\chi^2_{noFSI}$ & $\chi^2_{exFSI}$ & $\chi^2_{RFG}$ & $\chi^2_{SF}$ & $\chi^2_{GENIE}$ & $\chi^2_{noEl}$ \\
 \hline
 & \multicolumn{8}{|c|}{Shape-only} \\
  \hline
T2K reg. ($\delta p_T$) &  3.3 & 29.0 & 88.2 & 20.0 & 45.8 & 10.0 & 43.5 & 111.4 \\
T2K no reg. ($\delta p_T$) &  3.9 & 28.5 & 86.9 & 22.1 & 42.6 & 10.5 & 45.7 & 111.9 \\
\hline
T2K reg. ($\delta \alpha_T$) &  21.9 & 18.1 & 17.8 & 44.1 &  16.0 & 17.8 & 46.6 & 40.1 \\
T2K no reg. ($\delta \alpha_T$) &  21.2 & 17.9 & 19.0 & 38.1 & 18.0 & 18.9 & 39.4 & 39.4 \\
\hline
T2K reg. ($\delta \phi_T$) &  6.9 & 21.1 & 66.4 & 43.4 & 16.1 & 8.0 & 122.8 & 124.0 \\
T2K no reg. ($\delta \phi_T$) &  7.8 & 20.2 & 61.0 & 41.3 & 16.7 & 9.1 & 115.3 & 122.3 \\
\hline
 & \multicolumn{8}{|c|}{Full} \\
 \hline
T2K reg. ($\delta p_T$) &  31.4 & 62.1 & 371.9 & 36.3 & 129.6 & 21.8 & 72.4 & 153.6 \\
T2K no reg. ($\delta p_T$) &  32.7 & 62.7 & 370.1 & 37.4 & 123.1 & 22.9 & 72.7 & 149.3 \\
\hline
T2K reg. ($\delta \alpha_T$) &  60.5 & 34.9 & 107.5 & 65.3 & 51.9 & 29.1 & 68.3 & 50.8 \\
T2K no reg. ($\delta \alpha_T$) &  62.9 & 37.2 & 110.7 & 66.5 & 54.7 & 31.2 & 67.3 & 52.0 \\
\hline
T2K reg. ($\delta \phi_T$) &  36.9 & 48.8 & 303.5 & 64.2 & 61.2 & 17.8 & 170.5 & 165.4 \\
T2K no reg. ($\delta \phi_T$) &  39.1 & 49.7 & 300.5 & 63.8 & 62.5 & 19.0 & 167.8 & 159.0 \\
\hline
\end{tabular}
\caption{A summary of the $\chi^2$ calculated for comparisons of the full and shape-only T2K results with and without regularisation to the various models from NEUT and GENIE considered within this manuscript. All T2K results contain 8 bins.\label{tab:chi2regcomp}}
\end{table}

\end{document}